\documentclass[sigconf,anonymous=False,review=False]{acmart}
\newcommand\mymodel{FairNeg}

\usepackage{enumitem}
\usepackage{subfigure}
\usepackage[linesnumbered,ruled,vlined]{algorithm2e}
\usepackage{bm}
\usepackage{multirow}
\usepackage{fancyhdr,graphicx,amsmath}
\usepackage[T1]{fontenc}
\usepackage{flushend}
\usepackage{pifont}
\usepackage{listings}
\usepackage{color}
\definecolor{codegreen}{rgb}{0.3,0.5,0.0}
\AtBeginDocument{%
  \providecommand\BibTeX{{%
    \normalfont B\kern-0.5em{\scshape i\kern-0.25em b}\kern-0.8em\TeX}}}

\copyrightyear{2023}
\acmYear{2023}
\setcopyright{acmlicensed}
\acmConference[WWW '23]{Proceedings of the ACM Web Conference 2023}{May 1--5, 2023}{Austin, TX, USA}
\acmBooktitle{Proceedings of the ACM Web Conference 2023 (WWW '23), May 1--5, 2023, Austin, TX, USA}
\acmPrice{15.00}
\acmDOI{10.1145/3543507.3583355} 
\acmISBN{978-1-4503-9416-1/23/04}

\begin{document}
\title{Fairly Adaptive Negative Sampling for Recommendations}

\author{
    Xiao Chen$^{1,3}$, 
    Wenqi Fan$^{1*}$, 
    Jingfan Chen$^{1,3}$, 
    Haochen Liu$^2$ \\
    Zitao Liu$^5$,  
    Zhaoxiang Zhang$^{1,3,4*}$, 
    Qing Li$^1$\\ 
    {$^1$The Hong Kong Polytechnic University},
    {$^2$Michigan State University} \\ 
    {$^3$Center for Artificial Intelligence and Robotics, HKISI CAS} \\ 
    {$^4$Institute of Automation, Chinese Academy of Sciences} \\ 
    {$^5$Guangdong Institute of Smart Education, Jinan University}
}

\email{{shawn.chen,jingfan.chen}@connect.polyu.hk, wenqifan03@gmail.com, liuhaoc1@msu.edu}
\email{liuzitao@jnu.edu.cn, zhaoxiang.zhang@ia.ac.cn, csqli@comp.polyu.edu.hk}

\renewcommand{\shortauthors}{Chen and Fan, et al.}
\renewcommand{\authors}{Xiao Chen, Wenqi Fan, Jingfan Chen, Haochen Liu, Zitao Liu,  Zhaoxiang Zhang, Qing Li}
    
\begin{abstract}
Pairwise learning strategies  are prevalent for optimizing recommendation models on implicit feedback data, which usually learns user preference by discriminating between positive (i.e., clicked by a user) and negative items (i.e., obtained by negative sampling). However, the size of different item groups (specified by item attribute) is usually unevenly distributed. We empirically find that the commonly used uniform negative sampling strategy for pairwise algorithms (e.g., BPR) can inherit such data bias and oversample the majority item group as negative instances, severely countering group fairness on the item side. In this paper, we propose a \textbf{Fair}ly adaptive \textbf{Neg}ative sampling approach (\textbf{FairNeg}), which improves item group fairness via adaptively adjusting the group-level negative sampling distribution in the training process. In particular, it first perceives the model's unfairness status at each step and then adjusts the group-wise sampling distribution with an adaptive momentum update strategy for better facilitating fairness optimization. Moreover, a negative sampling distribution Mixup mechanism is proposed, which gracefully incorporates existing importance-aware sampling techniques intended for mining informative negative samples, thus allowing for achieving multiple optimization purposes. Extensive experiments on four public datasets show our proposed method's superiority in group fairness enhancement and fairness-utility tradeoff.
\end{abstract}

\settopmatter{printfolios=true} 
\thanks{* corresponding authors}
\vskip -0.30in
\maketitle

\begin{figure}[h]
    \centering
    \vskip -0.1in
    \includegraphics[width=0.85\columnwidth]{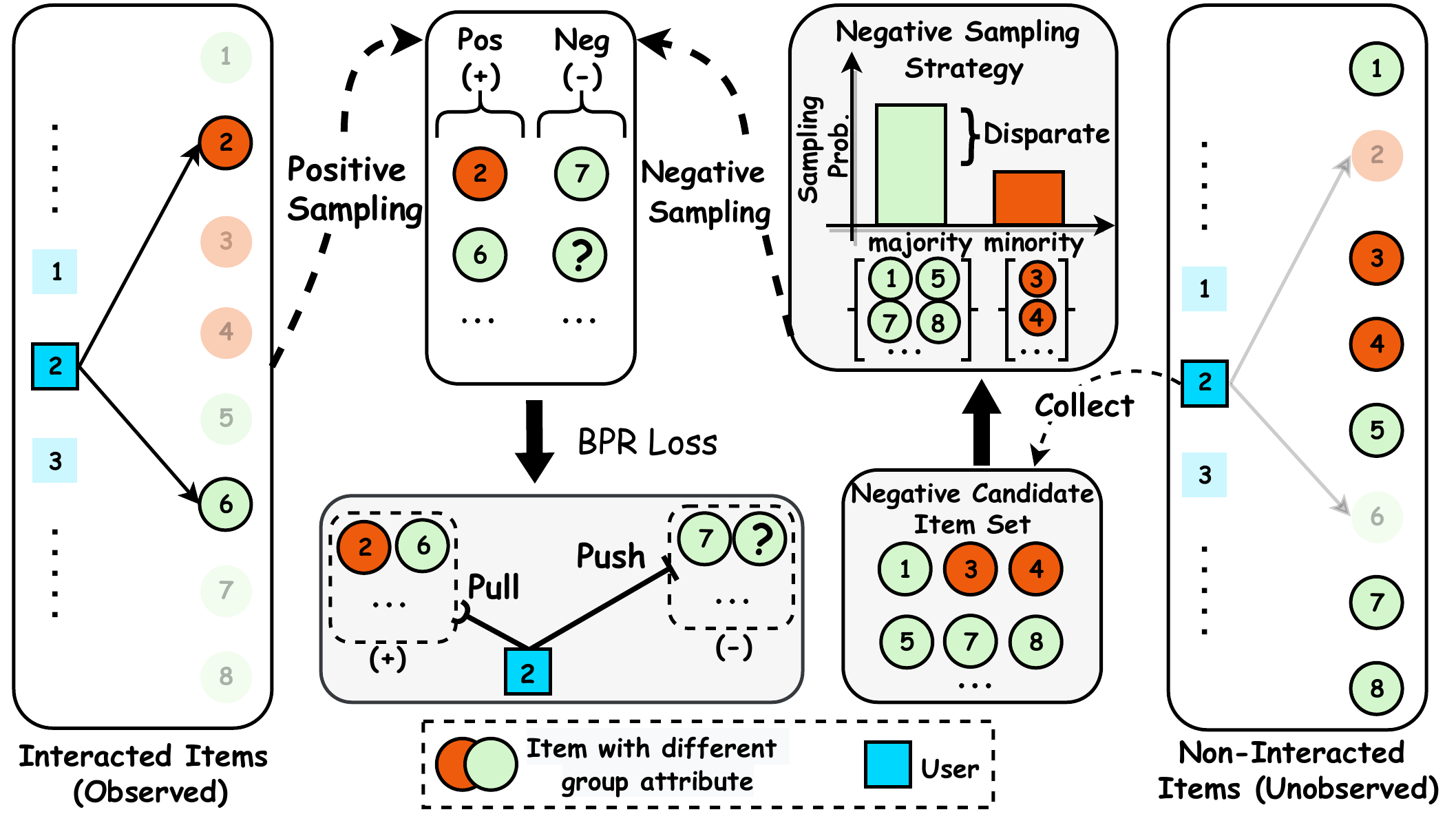}
    \vskip -0.10in
    \caption{
    Illustration of bias in Uniform Negative Item Sampling. 
    There are two groups of items  (determined by sensitive attributes, such as job payment or movie genres). Unobserved items in the majority group (in green color) have a higher probability of being sampled as negative items than that of the minority item group (in red color).}
    \label{fig:neg_sampling}
    \vskip -0.20in
\end{figure}

\section{Introduction}

Personalized recommender systems have been widely deployed to alleviate information overload issues~\cite{fan2019deep_dscf,fan2022comprehensive} by providing users with relevant items on various online services~\cite{fan2020graph, fan2022comprehensive, zhao2021autoloss,fan2021attacking}, such as online shopping, job matching, online education, etc. 
As the most representative techniques in recommender systems,  Collaborative Filtering (CF) methods~\cite{chen2022knowledge, fan2018deep} aim to learn recommendation models  by encoding user-item implicit feedback (e.g., user clicks and purchases) with a pairwise learning paradigm, e.g., Bayesian Personalized Ranking (BPR)~\cite{rendle2012bpr, fan2019deep_daso}.
In general, such a mainstream learning paradigm first conducts randomly negative item sampling to construct training data~\cite{he2020lightgcn,fan2022graph}, where it usually pairs a positive item (from observed user interactions) with a negative item (from unobserved interactions), and then encourages the predicted score of the positive item to be higher than that of the negative item.

However, such a mainstream learning paradigm is vulnerable to data bias~\cite{chen2020bias,liu2022self} on the item side, considering that the size of different item groups (determined by a specific item attribute) is highly imbalanced in the real world. 
For example, \emph{low-paying} jobs (majority group) occupy a larger portion than \emph{high-paying} jobs (minority group) in the job matching system, and the number of \emph{comedy} movies overpasses the \emph{Horror} movies for movie recommendations, both of which can be viewed as the inherent data bias on the item side~\cite{gomez2020characterizing}.
Moreover, to optimize recommendation models via the BPR loss, the prevalent negative sampling technique, i.e., Uniform Negative Sampling (\textbf{UNS}), is more likely to treat unobserved items in the majority group as negative. 

For instance, as illustrated in Figure~\ref{fig:neg_sampling}, items in the majority group (in green color) gain higher sampling probability as negative instances than that of the minority group (in red color) when learning for user $u_2$'s preference. 
As a result, in the training process, the majority item group will be pushed to obtain low (biased) prediction scores via the BPR loss, resulting in disparate recommendation performance (e.g., Recall) over different item groups.
In other words, such a uniform negative sampling strategy naturally inherits group-wise data bias on the item side, which misleads recommendation models into learning biased users' preferences and dampens the recommendation quality for items in the majority group.

\begin{figure}
\centering
\vskip -0.0510in
\subfigure[Data Distribution]{\label{fig:ds}\includegraphics[width=0.25\linewidth,height=0.3\linewidth]{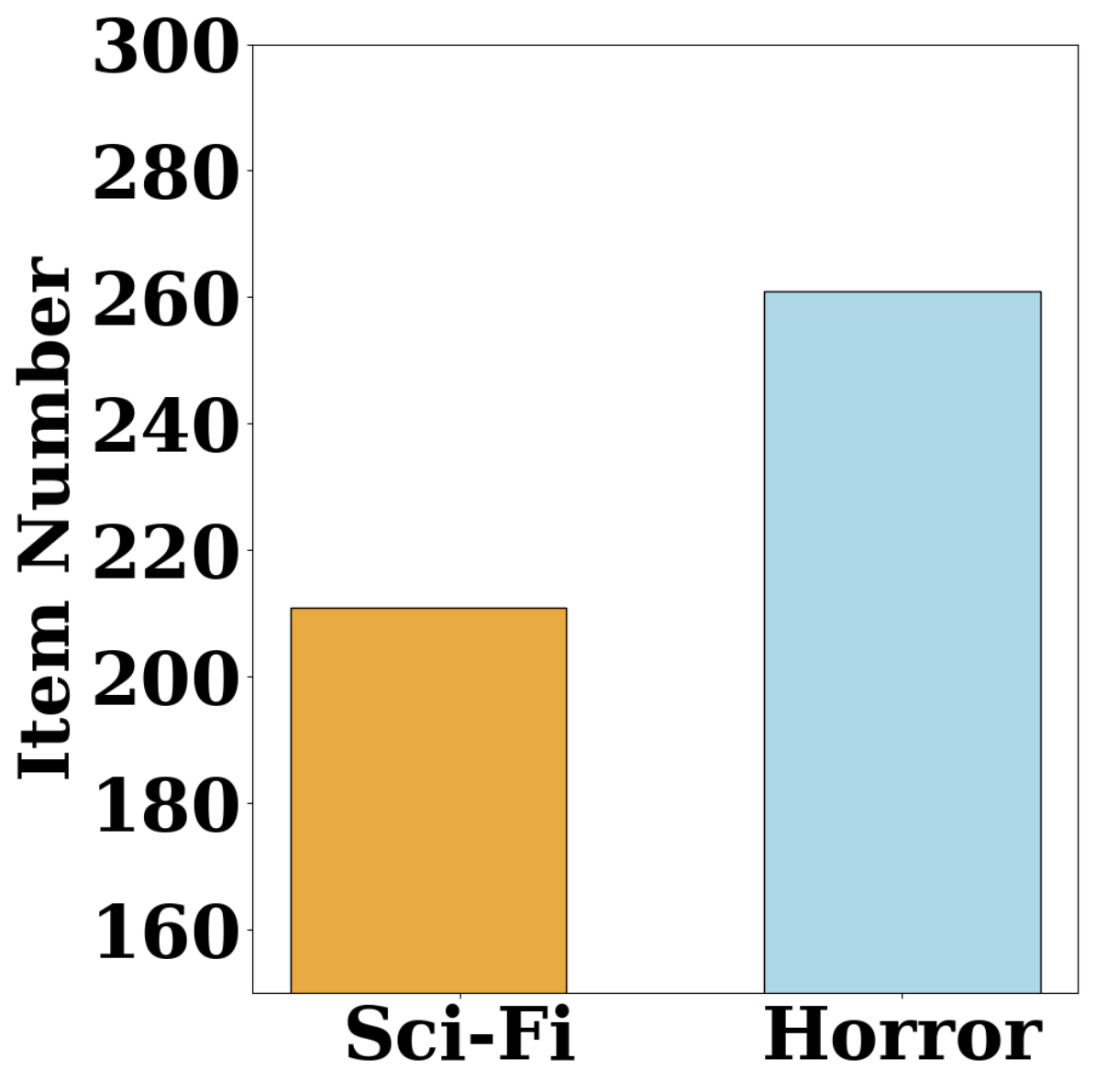}}
\subfigure[MF]{\label{fig:mf}\includegraphics[width=0.35\linewidth,height=0.3\linewidth]{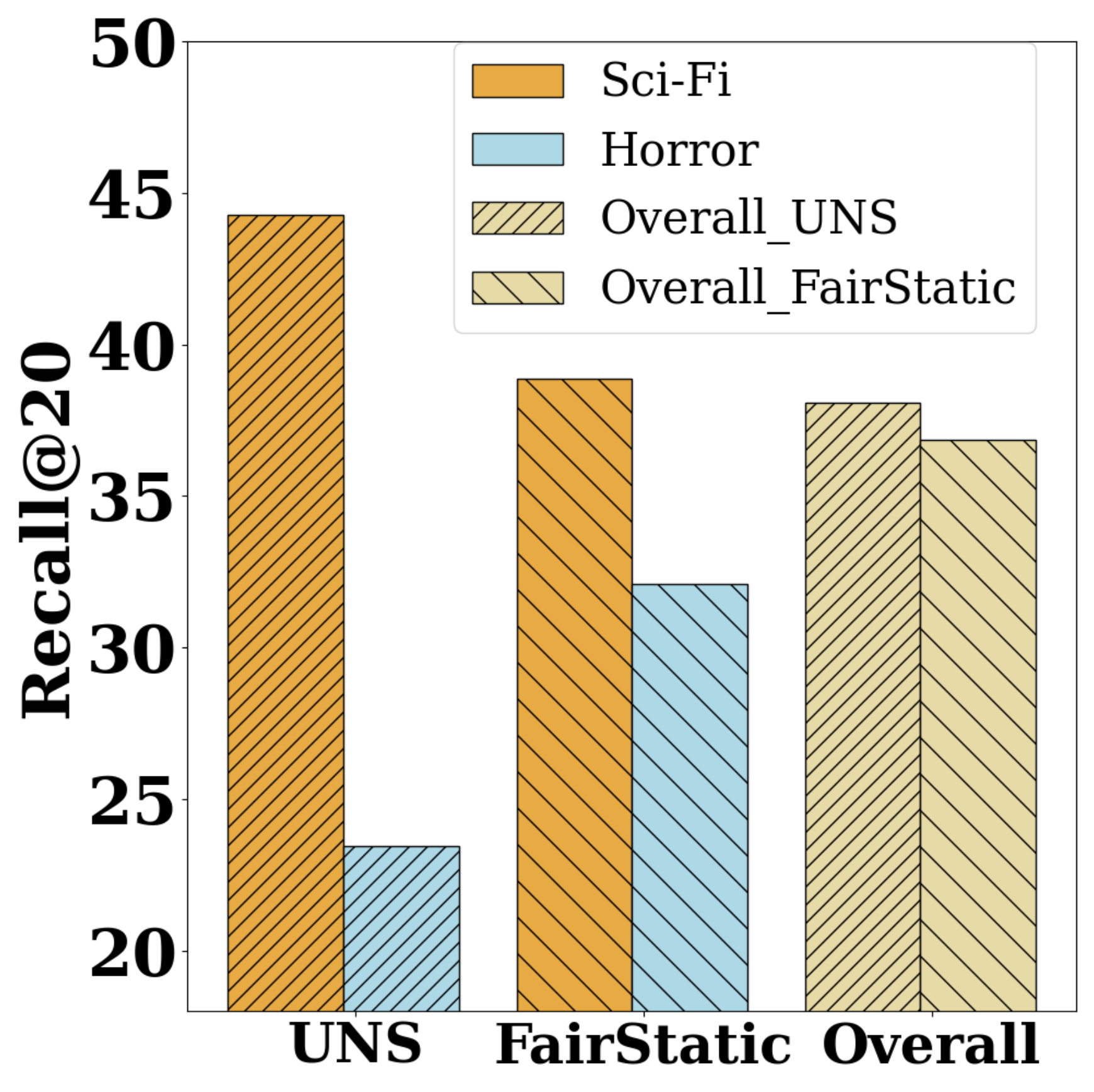}}
\subfigure[LightGCN]{\label{fig:lgn}\includegraphics[width=0.35\linewidth,height=0.3\linewidth]{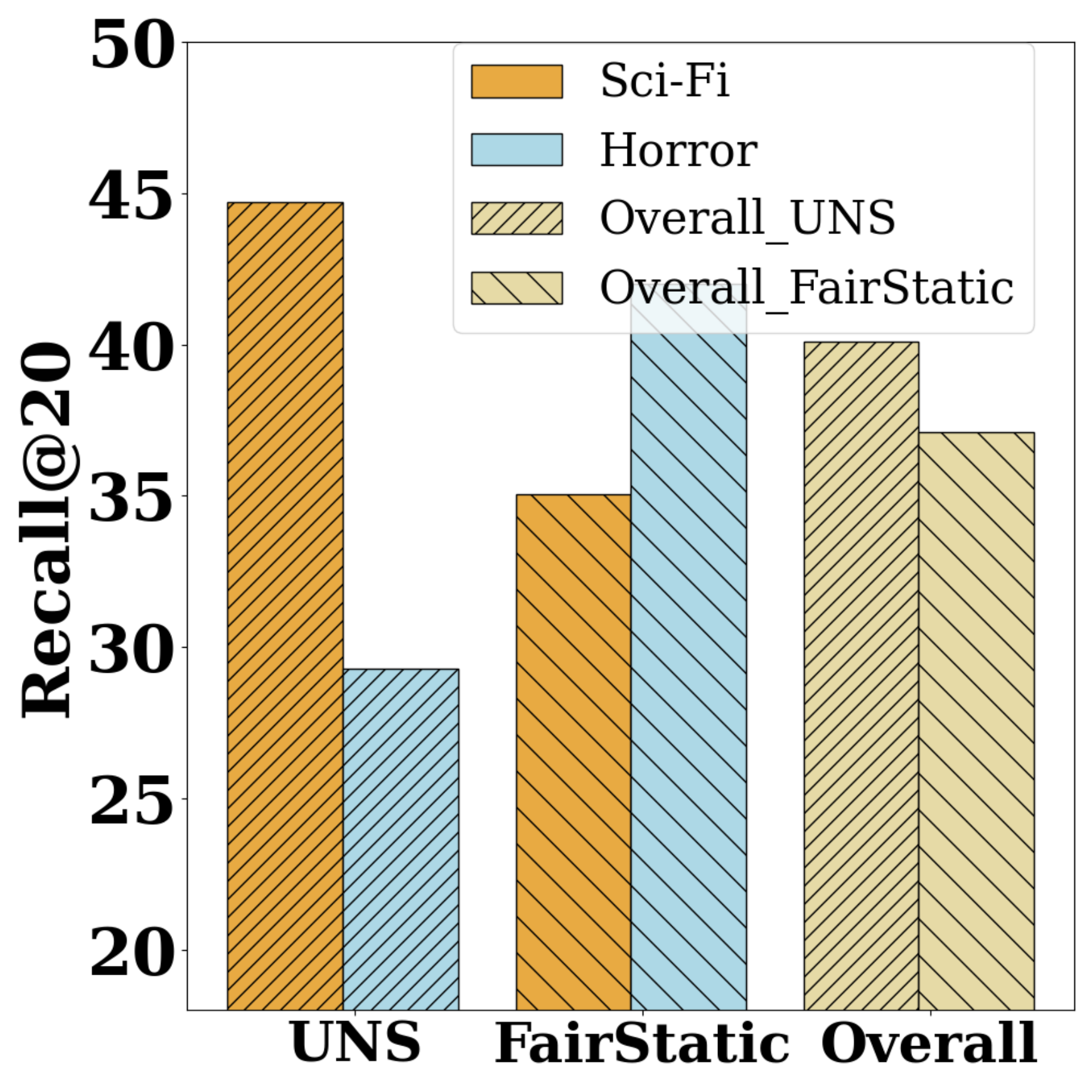}}
\vskip -0.10in
\caption{
(a) Distribution of two item groups on the ML1M-2 dataset. In (b) and (c), the $\text{Recall}@20$ performance of Sci-Fi and Horror movie groups are reported on two recommendation models with two different negative sampling strategies.}
\label{fig:intro_fig2}
\vskip -0.2in
\end{figure}

To address such undesired bias and unfairness caused by the negative sampling procedure, a straightforward solution is to adjust the negative sampling distribution by equalizing each item group's chance of being sampled as negative and positive instances, such that items in the majority group will not be oversampled as negative instances for the BPR optimization.
Here we call such a bias mitigation method as \textbf{FairStatic}. 
In order to verify the aforementioned unfair treatment over different item groups in UNS and its corresponding solution (i.e., FairStatic), we conduct preliminary studies to examine how these two sampling strategies (i.e., UNS v.s. FairStatic) perform on two recommendation methods: a classic Matrix Factorization model (MF)~\cite{rendle2012bpr} and an advanced Graph Neural Network-based model (LightGCN)~\cite{he2020lightgcn,Fan2023GenerativeDM}.  
For simplicity, we only consider the two-group scenario and work on the ML1M-2 dataset, whose movie items can be categorized into two genres (i.e., Sci-Fi and Horror). The detailed data description is shown in Experiments Section~\ref{sec:Experiments}, and the experimental results are shown in Figure~\ref{fig:intro_fig2}. 

It can be observed that MF/LightGCN+UNS methods exhibit a significant disparity in recommendation performance between the two item groups under $\text{Recall}@20$ metric, while MF/LightGCN+FairStatic could reduce the group-wise disparity at the cost of sacrificing overall recall performance.
From these observations, we conclude that adjusting the negative item sampling distribution can alleviate group unfairness on the item side. However, such naive adjustment based on the static (i.e., fixed in the whole training process) distribution greedily pursues the fairness objective, easily leading to a sub-optimal recommendation performance. 
Therefore, it is desirable to design an adaptive negative sampling method for achieving a better tradeoff between item group fairness and overall recommendation accuracy (utility).  

In this paper, we seek to address item-side group performance unfairness in recommendations from a novel perspective. Existing negative sampling strategies used in the pairwise learning paradigm usually neglect the item's group attribute~\cite{chen2017sampling, ding2020simplify}, which makes them 
vulnerable to item-side data bias and exhibit disparate group performance. Correspondingly, we propose a novel fairly adaptive negative sampling method (\textbf{FairNeg}), which can dynamically adjust the group-level negative sampling distribution based on the perceived performance disparity during the model training process. Moreover, FairNeg can be gracefully merged with importance-aware sampling techniques and yield a better utility-fairness tradeoff.
Our major contributions are summarized as follows: 
\begin{itemize}[leftmargin=*] 
\item  To the best of our knowledge, we are the first to investigate the bias and unfairness issues in recommendations from the perspective of negative sampling. We empirically show a principled connection between item-oriented group performance fairness and the group-level negative sampling distribution;
\item We propose a novel \textbf{Fair}ly adaptive \textbf{Neg}ative sampling framework (\textbf{FairNeg}) to mitigate item group unfairness in recommendations, where the group-level negative sampling distribution is updated adaptively via an outer optimizer, and the standard recommendation model is kept as the inner optimizer. These two optimizers altogether turn the training paradigm into a bi-level optimization framework, which enables easier-to-reconfigure implementation for fair recommendations;
\item Extensive experiments on different real-world datasets demonstrate the effectiveness of the proposed methods in significantly mitigating performance disparity among different item groups while maintaining the user-side overall recommendation quality. 
\end{itemize}
\begin{figure*}[t]
    \centering
    \vskip -0.10in
    \includegraphics[width=0.7\linewidth]{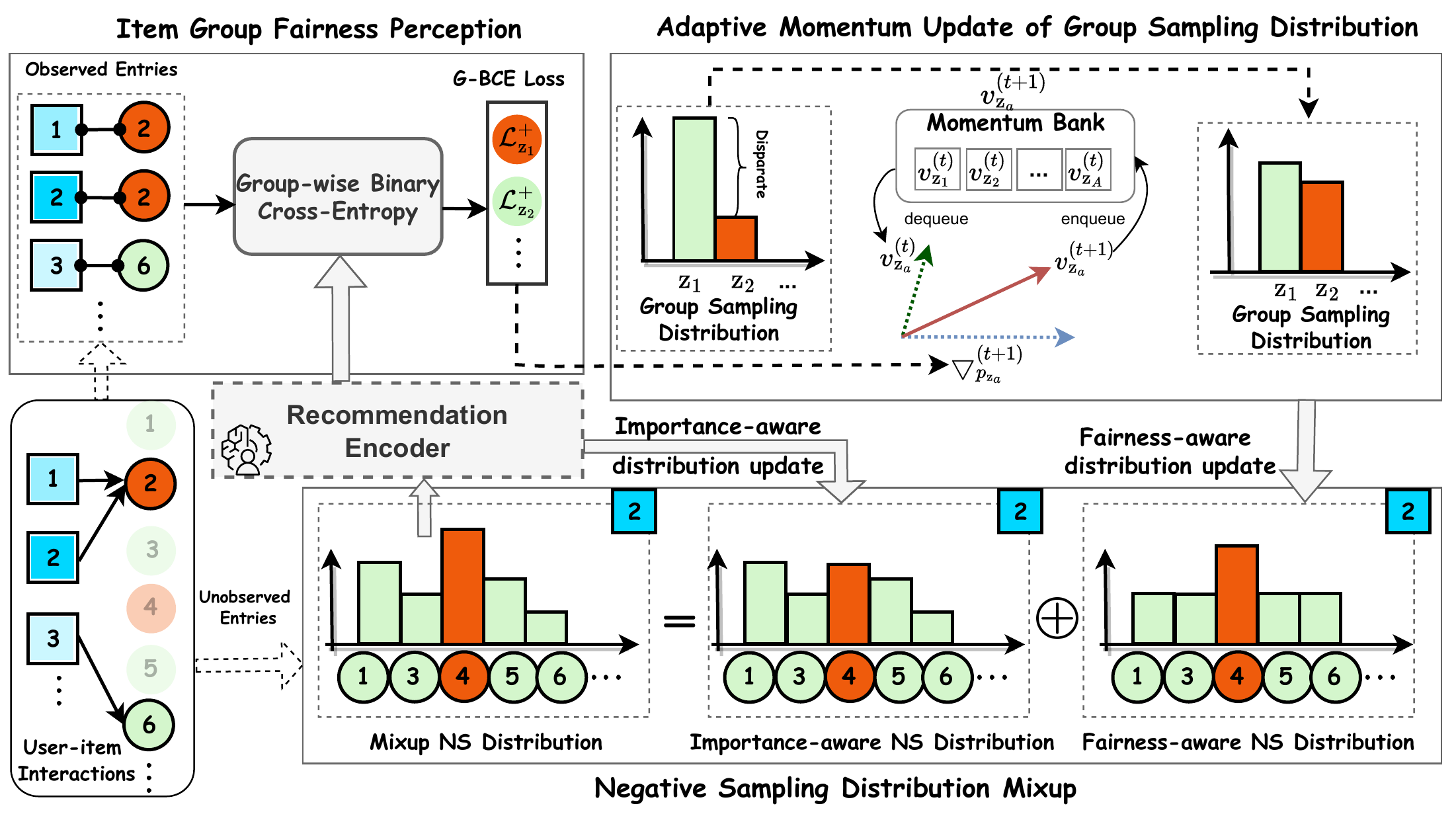}
    \vskip -0.10in
    \caption{Illustration of the proposed \mymodel{} framework. There are three components in this Fairly adaptive Negative sampling (FairNeg) framework: (i) Item Group Fairness Perception, (ii) Adaptive Momentum Update of Group Sampling Distribution, and (iii) Negative Sampling (NS) Probability Distribution Mixup.}
    \label{fig:my_framework}
    \vskip -0.150in
\end{figure*}

\section{Preliminaries}

In this section, we introduce some preliminary knowledge concerning the problem under study. We first present the notions related to the top-k item recommendation task~\cite{zhu2019improving} and the formulation of BPR loss. Then we introduce the definition and evaluation metric of item-oriented group performance fairness, which deserves our efforts in this paper.

\subsection{Personalized Ranking on Implicit feedback}

 \label{sec:algorithm}
Let $\mathcal{U}=\{u_{1},  \ldots, u_{m}\}$, $\mathcal{V}=\{v_{1}, \ldots, v_{n}\}$ be the sets of users and items respectively, 
and $\mathbf{Y} \in \mathbb{Y}^{m\times n}$ to represent the user-item historical (implicit) interaction matrix, where $y_{ui} = 1$ (in $\mathbf{Y}$) represents an observed interaction between user $u$ and item $i$, and 0 otherwise. 
In addition, we use $\mathcal{V}_{u}^{+}$ to denote the interacted (positive) item set of user $u$, and $\mathcal{V}_{u}^{-}= \mathcal{V} - \mathcal{V}_{u}^{+}$ to represent the un-interacted (negative) item set.
In this paper, we focus on the top-$K$ item recommendation task based on the implicit interactions $\mathbf{Y}$, which can be generally optimized by the pairwise BPR loss $\mathcal{L}_{\operatorname{BPR}}$ for learning model's parameters $\Theta$.
The main idea is to encourage the predicted scores on observed (positive) interactions to be higher than those of unobserved (negative) items, which can be formalized as follows:
\vskip -0.150in
\begin{equation}
\hspace*{-3.2mm}
\fontsize{8pt}{8pt}\selectfont
\begin{aligned}
\label{eq:bprloss}
    \min _{\Theta} \mathcal{L_{\operatorname{B P R}}}=-\sum_{u \in \mathcal{U}} \sum_{\substack{i \in  \mathcal{V}_{u}^{+}, j \in \mathcal{V}^-_u }} \ln \sigma\left(\hat{y}_{u, i}-\hat{y}_{u, j}\right), 
\hspace*{-5mm} 
\end{aligned}
\fontsize{10pt}{10pt}\selectfont
\end{equation}

where the negative item $j$ is sampled from the unobserved item set $\mathcal{V}^-_u$ of user $u$, 
$\hat{y}_{u, i}$ and $\hat{y}_{u, j}$ denote the predicted ranking scores generated by the recommendation model $f(\cdot)$ with parameters $\Theta$, and $\sigma$ is the sigmoid function. Here we ignore the regularization term for simplicity. 

\subsection{Problem Definition}
We assume a set of item groups with size A, $Z=\left\{\mathrm{z}_{1},\mathrm{z}_{2},...,\mathrm{z}_{A}\right\}$, specified by the item attribute (e.g., job payment, ethnicity, or movie genre). These groups are non-overlapping since each item belongs to one group. In this paper, we focus on enhancing \textbf{Item-oriented Group Performance Fairness} of recommendation models. 

In particular, to evaluate whether recommendation models treat different item groups equally, in the test stage, we can measure each group's recommendation performance using metrics such as $\text{Recall}@k$, which is an essential metric that reflects the coverage of truly relevant items in the top-k recommendation results for one item group. 
Then, the disparity of group-wise $\text{Recall}@k$ performance
(abbreviated as $\text{Recall-Disp}@k$) can be calculated following the definition of previous work~\cite{zhu2020measuring}:
\begin{align}
\small{
\label{eq:Recall-Disp}
\text{Recall-Disp}@k= \frac{\text{std}(\text{Recall}@k|z=\mathrm{z}_{1},..., \text{Recall}@k|z=\mathrm{z}_{A})}
{{\text{mean}(\text{Recall}@k|z=\mathrm{z}_{1},..., \text{Recall}@k|z=\mathrm{z}_{A})}},
}
\end{align}
where $\text{std}(\cdot)$ is the standard deviation function and $\text{mean}(\cdot)$ calculates the mean average value. Such relative standard deviation can be computed to reflect the group-wise performance disparity. The item-oriented fairness (indicated by the lower value of $\text{Recall-Disp}@k$) is supposed to be fulfilled in general item recommendation systems so that all items, regardless of their group attributes, have the same chance of being exposed to users who truly like them, and no user needs are ostensibly ignored.
\section{The Proposed Method}
In this section, we will first give an overview of the proposed method \mymodel{}, then describe each of its components in detail and finally illustrate the designed optimization algorithm for item group fairness.

\subsection{An Overview of the Proposed Framework}
Since the empirical risk minimization (i.e., BPR loss) does not consider the fairness criterion, the recommendation model usually exhibits severe unfairness issues. To address such limitations without explicitly changing the model architecture or training data, 
we design a fairly adaptive negative sampling framework \mymodel{}, as shown in Figure~\ref{fig:my_framework}, which adjusts each item group's negative sampling probability for fairness considerations and incorporates importance-aware technique for enhancing negative samples' informativeness.
The proposed \mymodel{} method consists of three main components: (i) Item Group Fairness Perception, (ii) Adaptive Momentum Update of Group Sampling Distribution, and (iii) Negative Sampling Distribution Mixup. The first component, Item Group Fairness Perception, is designed to perceive unfairness (i.e., recall disparity) among item groups during the training process. Next, with the perceived group performance disparity, the second component (i.e., Adaptive Momentum Update of Group Sampling Distribution) adjusts \textit{each group's negative sampling probability} accordingly, intending to equalize all groups' recall performance. At last, the third component (i.e., Negative Sampling Distribution Mixup) further incorporates an \textit{importance-aware} sampling probability with a mixup mechanism. In this way, the group fairness optimization and the informativeness of negative samples can be jointly considered.

\subsection{Item Group Fairness Perception} 
\label{sec:3.2}
To achieve fair training from the perspective of adaptive negative sampling, we first need to perceive the group-wise performance disparity~\cite{hardt2016equality,mcnamara2019equalized,roh2021fairbatch}, which provides guidance for adjusting the group-level negative sampling distribution in the next step. However, the fairness evaluation metric (i.e., $\text{Recall-Disp}@k$) mentioned above cannot be used here \textit{directly} in the training phase since it is non-differentiable.
We hereby propose a Group-wise Binary Cross-Entropy~(G-BCE) loss as a proxy to measure each item group's recall performance. It only considers observed interactions (viewed as positive) in the training set and calculates the group-wise average classification loss as an approximation of group recall~\cite{hardt2016equality}. 
Next, we will introduce the details of G-BCE loss, followed by the analysis revealing that mitigating the disparity of \textit{multiple} groups' G-BCE losses at the training stage actually optimizes item-oriented group fairness.

More specifically, G-BCE loss measures the average difference between the predicted preference scores $\hat{y}_{u,v}$ and the truly observations $y_{u,v}$ in a specific item group $\mathrm{z}_{a}$:
\vskip -0.150in
\begin{equation}
\hspace*{-3.2mm}
\fontsize{8pt}{8pt}\selectfont
\begin{aligned}
\label{eq:gbce-loss}
\nonumber \mathcal{L}_{\mathrm{z}_{a}}^{\text{+}}=&-\frac{1}{|Y^{\text{+}}_{\mathrm{z}_{a}}|}\textstyle\sum_{(u,v)\in Y^{\text{+}}_{\mathrm{z}_{a}}}\Bigl( \underbrace{ y_{u,v} \cdot \log \sigma\left(\hat{y}_{u,v}\right) }_{\text{Positive Part}} \\
\nonumber &+\underbrace{ \left(1-y_{u,v}\right) \cdot \log \left(1- \sigma\left(\hat{y}_{u,v}\right)\right)}_{\text{Negative Part}} \Bigl)\\
=&-\frac{1}{|Y^{\text{+}}_{\mathrm{z}_{a}}|}\textstyle\sum_{(u,v)\in Y^{\text{+}}_{\mathrm{z}_{a}}} \log \sigma\left(\hat{y}_{u,v}\right),
\hspace*{-5mm} 
\end{aligned}
\fontsize{10pt}{10pt}\selectfont
\end{equation}
\noindent where $Y^{\text{+}}_{\mathrm{z}_{a}}$ denotes the set of all true observations for item group $\mathrm{z}_{a}$, $\hat{y}_{u,v}$ denotes the predicted preference score of user $u$ to item $v$, and $\sigma$ is the sigmoid function that normalizes the preference score to $\left ( 0,1 \right ) $. 
Note that the negative part in Equation~\ref{eq:gbce-loss} can be eliminated, since we only consider observed entries (referring to $y_{u,v}$ equals to 1) and $\log\left(\sigma\left(\hat{y}_{u,v}\right)\right)$ is equivalent to $\log\left(p\left ( \sigma\left(\hat{y}_{u,v}\right) = 1 \right ) \right)$.
As a result, $\mathcal{L}_{\mathrm{z}_{a}}^{\text{+}}=-\frac{1}{|Y^{\text{+}}_{\mathrm{z}_{a}}|}\textstyle\sum_{(u,v)\in Y^{\text{+}}_{\mathrm{z}_{a}}} \log \sigma\left(\hat{y}_{u,v}\right)$ can be viewed as an unbiased estimation of $-\log p(\hat{y}=1 \mid y=1, \mathrm{z}=\mathrm{z}_{a})$, which further approximates $-\log \left ( \text{Recall} \mid \mathrm{z}=\mathrm{z}_{a} \right)$.

Taking the two groups scenario as an example, if $\mathcal{L}_{\mathrm{z}_{1}}^{\text{+}} > \mathcal{L}_{\mathrm{z}_{2}}^{\text{+}}$ in the training phase, it indicates that the item group $\mathrm{z}_{1}$ (disadvantaged group) obtains lower recall performance. In other words, the recommendation model assigns a lower recommendation probability to items in group $\mathrm{z}_{1}$, given that users truly like these items. Next, we focus on minimizing the disparity of G-BCE loss (Group-wise Recall) during training, which is equivalent to optimizing toward item-oriented group fairness (as previously defined).

\subsection{Adaptive Momentum Update of Group Sampling Distribution}

In this work, we propose to achieve group fairness from a novel perspective -- negative sampling, which requires no change of model architecture or the training data. 
We attribute the dampening of the majority item group's recall performance (as shown in Figure~\ref{fig:intro_fig2}) to the overwhelmed negative gradients, which is caused by the oversampling of being negatives of this group.

Concretely, when different groups' item size is fairly imbalanced, the majority item group is likely to be assigned with a relatively higher negative sampling probability. Note that in BPR loss (as shown in Equation~\ref{eq:bprloss}), the paired negative item will receive discouraging gradients. Thus, the accumulated negative gradients of the majority group will make them obtain low predicted relevance scores, regardless of users' true preference on this group.

\begin{figure}[t]
\vskip -0.10in
    \centering
    \subfigure[Advantaged Item Group]{\label{fig:ds}\includegraphics[width=0.47\linewidth,height=0.36\linewidth]{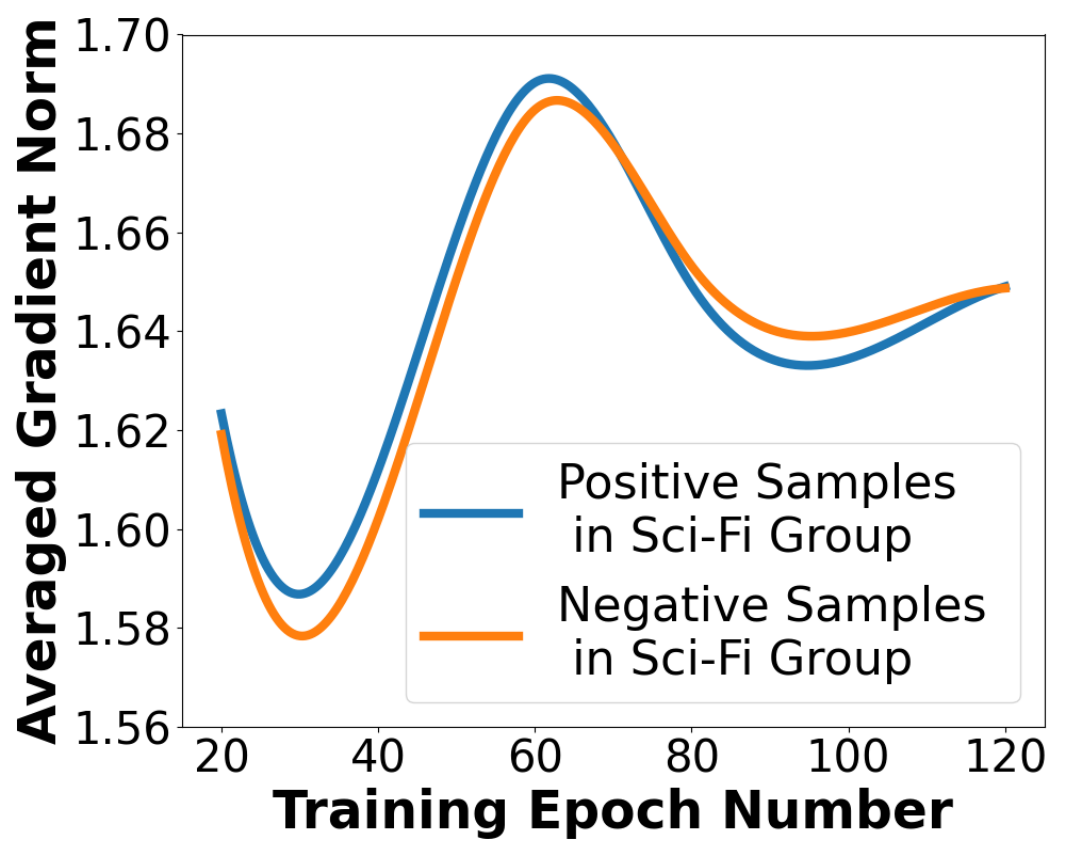}}
    \subfigure[Disadvantaged Item Group]{\label{fig:mf}\includegraphics[width=0.48\linewidth,height=0.36\linewidth]{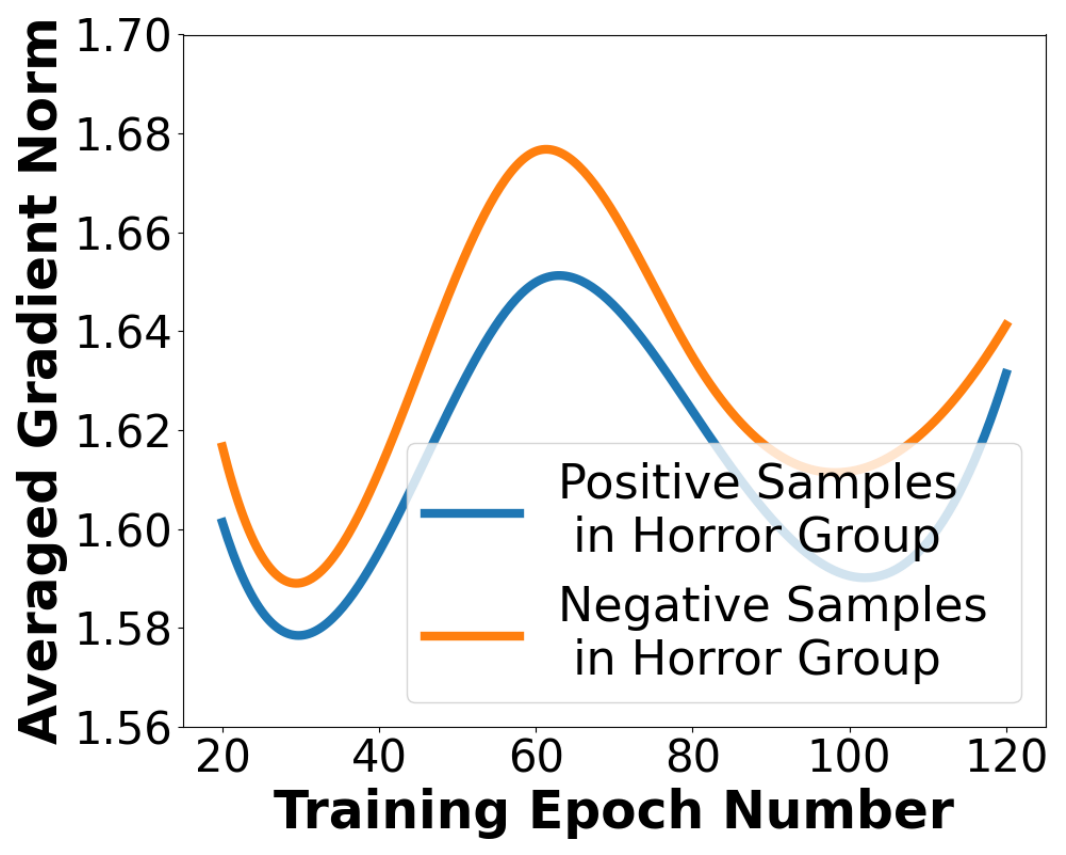}}
    \vskip -0.130in
    \caption{ The overall gradient analysis on positive and negative samples in two item groups on the ML1M-2 dataset (as described in Table~\ref{tab: datasets}) and the backbone is a Matrix Factorization model. We collect the averaged L2 norm of the gradient of weights~\cite{tan2020equalization} in the item embedding layer at each epoch until model convergence.}
    \label{fig:method_gradient}
    \vskip -0.15in
\end{figure}

To verify this, we conduct an empirical analysis to show the \textit{overwhelming negative gradients} on the majority item group during the optimization process, as illustrated in Figure~\ref{fig:method_gradient}. It is clear that the disadvantaged group (i.e., Horror group with worse recall performance) receives larger gradients from negative samples due to the high negative sampling probability of this group. 

Based on the above analysis, we are inspired to design the adaptive negative sampling strategy in two consequent steps, where the (dis)advantaged group is first identified based on the perceived G-BCE losses (as illustrated in Section~\ref{sec:3.2}), and then the negative sampling probability of this group will be updated so that the disadvantaged group is less overwhelmed by negative gradients.

Specifically, we assign each item group a negative sampling probability and update it via (1) \textit{Group-wise Gradient Calculation}, which calculates the current step's gradient of each group's sampling probability, and (2) \textit{Adaptive Momentum Update}, which utilizes the auxiliary knowledge from historical gradients to alleviate optimization oscillations.

\noindent{\textbf{Group-wise Gradient Calculation}}. 
Considering items with a multi-valued sensitive attribute (i.e., ${A}\ge2$), we define the group-wise negative sampling probability distribution as:
\begin{align}
 {\bm p}:&=\left ( p_{\mathrm{z}_{1}},p_{\mathrm{z}_{2}},\cdots,p_{\mathrm{z}_{A}}\right ), \text{where} \sum_{p_{\mathrm{z}_{a}} \in {\bm p}} p_{\mathrm{z}_{a}} = 1.
\end{align}

To narrow the performance gap among different groups, as illustrated in the previous analysis, we consistently adjust each group's sampling probability $\bm p$ according to its deviation from the averaged G-BCE loss at each step:
\begin{align}
\label{eq: grp_prob_1}
 \bigtriangledown_{p_{\mathrm{z}_{a}}}^{\left ( t \right )}:= \mathcal{L}_{\mathrm{z}_{a}}^{+^{\left ( t \right )}}-\frac{1}{\left | A \right | }{\textstyle\sum_{\mathrm{z} \in Z}}\mathcal{L}_{\mathrm{z}}^{+^{\left ( t \right )}},
\end{align}
where $\bigtriangledown_{p_{\mathrm{z}_{a}}}^{\left ( t \right )}$ denotes the gradient of $p_{\mathrm{z}_{a}}$ at step $t$, $\mathcal{L}_{\mathrm{z}_{a}}^{+^{\left ( t \right )}}$ denotes the G-BCE loss of group $\mathrm{z}_{a}$ at step $t$, and $\frac{1}{\left | A \right | }{\textstyle\sum_{\mathrm{z} \in Z}}\mathcal{L}_{\mathrm{z}}^{+^{\left ( t \right )}}$ denotes the step $t$'s averaged G-BCE loss. Intuitively, a disadvantaged item group with a relatively large G-BCE loss (low recall performance) will receive a negative-valued update vector, which down-weights its negative sampling probability during the next step's training. Such an adaptive down-weighting scheme prevents the disadvantaged group from being overwhelmed by negative gradients and is therefore expected to boost the group's recall performance.

\noindent{\textbf{Adaptive Momentum Update}.} 
Furthermore, to alleviate the adverse influence of gradient noise~\cite{sutskever2013importance} and instability of gradient direction, we design an adaptive momentum update strategy to produce the sampling probability of each item group at step $t+1$, which can be summarized into two steps:

\vskip -0.15in
\begin{align}
\label{eq:grp_prob_2}
       \nonumber v_{\mathrm{z}_{a}}^{\left ( t + 1 \right ) } &= \gamma v_{\mathrm{z}_{a}}^{\left ( t \right ) }  
        + \alpha \cdot \bigtriangledown_{p_{\mathrm{z}_{a}}}^{\left ( t+1\right )}, \\
 p_{\mathrm{z}_{a}}^{(t+1)} &= p_{\mathrm{z}_{a}}^{(t)}- v_{\mathrm{z}_{a}}^{\left ( t + 1 \right ) },
\end{align}
\vskip -0.10in

where current step's momentum update vector $v_{\mathrm{z}_{a}}^{\left ( t+1 \right ) }$ is a combination of last step's momentum vector $v_{\mathrm{z}_{a}}^{\left ( t \right ) }$ and current step's gradient $\bigtriangledown_{p_{\mathrm{z}_{a}}}^{\left ( t+1\right )}$, $\gamma \in\left [ 0,1 \right ]$ is the momentum coefficient, $\alpha$ is the learning rate; group-wise sampling probability $p_{\mathrm{z}_{a}}^{(t+1)}$ at step $t+1$ is then updated with the momentum update vector $v_{\mathrm{z}_{a}}^{\left ( t+1 \right ) }$.
Here we maintain the \emph{momentum bank}~\cite{he2020momentum} as a queue of momentum update vectors: all groups' corresponding update vectors at the current step are enqueued, and the previous ones are dequeued. Intuitively, with the adaptive momentum update strategy, if the current epoch's gradient is in the same direction as the previous epoch, the updated value will be larger and helps to achieve faster convergence. By contrast, if the gradient direction conflicts, the updated value will be reduced, which facilitates smoothing out the variations and stable learning~\cite{cutkosky2020momentum}.

 \subsection{Negative Sampling Distribution Mixup}
Though updating group sampling distribution adaptively ensures that item group fairness can be jointly optimized during training,  the informativeness difference of items within the same group is neglected due to the equal-share of sampling probability.
To be more specific, when drawing a negative sample from the unobserved item set $\mathcal{V}^-_u$ for user $u$ only following the updated group sampling distribution, for a certain candidate item $j$ from the group $\mathrm{z}_{a}$, its fairness-aware sampling probability can be computed as:
\vskip -0.10in
\begin{equation}
\label{eq:beta_fair}
\hspace*{-3.2mm}
\fontsize{8pt}{8pt}\selectfont
    \begin{aligned}
    p_{j}^{fair} = {\frac {p_{\mathrm{z}_{a}}}{\textstyle \sum_{i \in \left | \mathcal{V}^-_u \right |}G_{z_{a}}\left ( i \right ) } },
    \hspace*{-5mm} 
    \end{aligned}
\fontsize{10pt}{10pt}\selectfont
\end{equation}

where $p_{\mathrm{z}_{a}}$ is the group-wise sampling probability, and $G_{\mathrm{z}_{a}}\left(i\right) $ is a indicator function to identify whether item $i$ in the negative candidate item set (unobserved) belongs to group $\mathrm{z}_{a}$. The indicator function returns 1 when item $j$'s group attribute is $\mathrm{z}_{a}$, otherwise 0.
It can be seen that, in the candidate item set of user $u$, items from the same sensitive group share equal sampling probability. Nonetheless, such a sampling strategy tends to sample negative items that are easier to distinguish from positive samples and inhibit improvement of user and item feature discriminativeness.
 
To address this limitation, we propose a Negative Sampling Distribution Mixup mechanism, which incorporate an importance-aware negative sampling strategy. Such a strategy is built upon the dynamic negative sampling method for BPR~\cite{zhang2013dns}. The importance of a candidate item refers to its relevance to the user. Formally, the importance-aware negative sampling probability of candidate item $j$ for the user $u$ can be calculated as follows:
\vskip -0.10in
\begin{equation}
\label{eq:beta_importance}
\hspace*{-3.2mm}
\fontsize{8pt}{8pt}\selectfont
    \begin{aligned}
    p_{j}^{imp} = \frac{\exp \left ( s_{u,j}/{\tau} \right ) }{{{\textstyle \sum_{i \in \left | \mathcal{V}^-_u \right |} \exp \left ( s_{u,i}/{\tau} \right ) }} }, 
    \end{aligned}
\fontsize{10pt}{10pt}\selectfont
\end{equation}

where $s_{u,j}$ is the user-item relevance score calculated by the recommendation model, and $\tau$ is a temperature parameter in the softmax function. The goal is to pick the high-scored item with a higher probability for user $u$, which facilitates learning more discriminative feature representations. 

Now that the importance-aware negative sampling distribution ($\bm p^{\text{imp}}$) implicitly improves recommendation accuracy by strengthening feature representations, and the fairness-aware negative sampling distribution ($\bm p^{\text{fair}}$) aids item-oriented group fairness. 
To incorporate them simultaneously, we introduce a mixup hyper-parameter $\beta$, which controls the strength of fairness optimization. The final sampling probability $p_{j}$ of candidate item $j$ for user $u$ can be formulated as: 
\vskip -0.15in
\begin{equation}
\hspace*{-3.2mm}
\fontsize{8pt}{8pt}\selectfont
\begin{aligned}
\label{eq:beta}
p_{j} = \beta \cdot p_{j}^{\text{fair}} + (1 - \beta) p_{j}^{\text{imp}}.
\hspace*{-5mm} 
\end{aligned}
\fontsize{10pt}{10pt}\selectfont
\end{equation}

With the new mixup negative sampling distribution computed for each user, the recommendation model training considers both item group fairness and user's preference learning. 

\subsection{Optimization for Item Group Fairness}

 We formulate the optimization of \mymodel{} as a  bi-level problem, where the optimization of the group-wise negative sampling distribution is nested within the recommendation model parameters optimization.
 This can be solved by alternatively updating the fairness-aware negative group sampling distribution and recommendation model parameters~\cite{talbi2013taxonomy}.
 More specifically, the negative sampling distribution is learned with the outer optimizer, while a personalized ranking algorithm (i.e., BPR) is kept as the inner optimizer for updating user and item representations.

Mathematically, such bi-level optimization can be formulated as follows:
\begin{small}
\vskip -0.150in
\begin{align}
\nonumber \bm{p}^{*} &=\underset{\bm p}{\arg \min }\mathcal{L}_{\text{Recall-Disp}}(\bm{\Theta}_{\bm{p}}):=\sum_{\mathrm{z_{a}} \in Z}\left | \mathcal{L}_{\mathrm{z}_{a}}^{+}-\frac{1}{\left | A \right | }{\textstyle\sum_{\mathrm{z} \in Z}}\mathcal{L}_{\mathrm{z}}^{+} \right |,  
\\
\bm{\Theta}_{p}^{*} &=\underset{\bm{\Theta}}{\arg \min } \mathcal{L}_{\operatorname{utility}}(\bm{\Theta},\bm{p}) := -\sum_{u \in \mathcal{U}} \sum_{\substack{i \in  \mathcal{V}_{u}^{+}, j \in \mathcal{V}^-_u }}  {\mathcal{L}_{\operatorname{BPR}}}\left ( u,i,j; \bm{\Theta},\bm{p}\right ),
\end{align}  
\end{small}

where $\bm{\Theta}$ is the recommendation model parameters, and $\bm{p}$ is the group-wise negative sampling distribution. ${L}_{\text{Recall-Disp}}$ represents the disparity of G-BCE loss, which is the sum of each item group's deviation from the macro-average performance. 

The optimization algorithm is presented in Algorithm~\ref{algo:algorithm_bilevel}. First, we initialize group-wise negative sampling distribution $\bm{p}$ as the way of FairStatic, where each group's sampling probability equals its proportion in the overall user-item positive interactions; recommendation model parameters are initialized with Xavier~\cite{glorot2010understanding} (line 1). Next, with $\bm{p}$ fixed, we conduct pairwise learning and iteratively update the recommendation model parameters $\bm{\Theta}$ using BPR loss for one epoch. The negative items in the training pairs are sampled based on a mixup distribution, considering both item importance and group fairness (lines 2-7). Then, we fix the model parameters $\bm{\Theta}$ and adjust each group's sampling probability to reduce their recall disparity (lines 8-9). The iteration will stop early when the recommendation performance does not improve for consecutive $p$ epochs (line 10).

\begin{algorithm}[h]
\caption{\textbf{Bi-level optimization of \mymodel{}}}
\label{algo:algorithm_bilevel}
\SetKwData{}{}
\KwIn{Train data ($\mathcal{U}$, $\mathcal{V}$, $\mathbf{Y}$), Epochs E}  \KwOut{Optimized recommendation model parameter $\bm{\theta^{*}}$} 
Initialize model parameters $\bm{\theta}$ and  group-wise negative sampling distribution ${\bm p}:=\left ( p_{\mathrm{z}_{1}},p_{\mathrm{z}_{2}},\cdots,p_{\mathrm{z}_{A}}\right )$\;
\For{$each \, epoch$}{
    \For{ $each \,$ $\left(u, i\right)\in Y^{+}$ }{
    Obtain the candidate item set $\mathcal{V}_{u}^{+}$ for negative sampling\;
    Calculate each candidate item's mixup sampling probability $p_{j}$ according to Equation~\ref{eq:beta_fair},~\ref{eq:beta_importance},~\ref{eq:beta}\;
    Sample a negative item $j$ according to the mixup sampling distribution\;
    Conduct pairwise learning and update model parameters $\bm{\theta}$ with $(u,i,j)$ based on equation~\ref{eq:bprloss}\;
    }
    Calculate G-BCE loss for each item group $\mathrm{z}_{a}$\;
    Update each group's sampling probability $p_{\mathrm{z}_{a}}$ based on Equation~\ref{eq: grp_prob_1},~\ref{eq:grp_prob_2}\;
    Early stop when the model performance on validation set doesn't improve for consecutive $p$ epochs;
} 
return;
\end{algorithm}
\vskip -0.10in

\section{Experiment}\label{sec:Experiments}

In this section, we conduct extensive experiments on four real-world datasets to demonstrate the effectiveness of our proposed \mymodel{}. 

\subsection{Experiment Settings}
\subsubsection{Datasets}
The experiments are performed on publicly available datasets -- MovieLens 1M (\textbf{ML1M})~\cite{harper2015movielens} and \textbf{Amazon} product reviews~\cite{mcauley2015image}. The ML1M dataset consists of movie ratings in different genres, and we treat all ratings as positive feedback. The Amazon dataset includes the historical product reviews collected from the Amazon e-commerce platform, and we view all user purchase behaviors as positive feedback.
Considering that some comparison methods can only work for the dataset with binary item attributes, we extract two subsets from each dataset, containing either a binary-valued or a multi-valued item attribute.
Specifically, we extract ratings with four genres (`Sci-Fi', `Adventure', `Children', and `Horror') to constitute our \textbf{ML1M-4} dataset, and keep the product reviews with four types (`Grocery', `Office', `Pet', and `Tool') to constitute our \textbf{Amazon-4} dataset.
Moreover, we create \textbf{ML1M-2} (`Sci-Fi' vs. `Horror') and \textbf{Amazon-2} (`Grocery' vs. `Tool') datasets by keeping the most popular (denoted by the average feedback number of this group) and least popular two groups. 
The statistics of the four datasets are detailed in Table~\ref{tab: datasets}, and the detailed group-wise statistics (i.e., feedback and item number) are illustrated in the Appendix Section~\ref{sec:statistics}. All datasets are randomly split into 60\%, 20\%, and 20\% to constitute the training, validation, and test sets. 

\begin{table}[t]
\centering
\vskip -0.10in
\caption{Characteristics of the datasets.}
\vskip -0.10in
\label{tab: datasets}
\scalebox{0.90}
{
\begin{tabular}{l|llll}
\hline
\hline
         & Users & Items & Ratings & Density \\ \hline
ML1M-2   & 4,680  &  472   & 194,610  & 8.81\%  \\
ML1M-4   & 5,120  &  685   & 242,419  & 6.91\%  \\
Amazon-2 & 3,845  & 2,487  &  84,656  & 0.89\%  \\
Amazon-4 & 3,954  & 3,061  & 129,433  & 1.07\%  \\ \hline
\hline
\end{tabular}
}
\vskip -0.20in
\end{table}

\subsubsection{{Evaluation Metrics}}

We evaluate the recommendation utility and the item group fairness in the experiments. For group fairness evaluation, we stratify the items by groups based on the item attribute, then compute \textbf{Recall@k} for each group based on the Top-$k$ recommendation results. Afterward, we report three group fairness metrics as in the literature~\cite{lahoti2020fairness}: (i) \textbf{Recall-Disp}: the Recall disparity among all groups, and the formula is described in Equation~\ref{eq:Recall-Disp}. (ii) $\textbf{Recall-Min}$: the minimum Recall among all groups. (iii) $\textbf{Recall-Avg}$: the macro-average Recall over all groups.
Here, the lower $\text{Recall-Disp}$ value represents better group fairness, and the higher values for the latter two metrics are better. For the recommendation utility evaluation on the user side, we employ standard Top-$k$ ranking metrics, including \textbf{Precision (P@k)}, \textbf{Recall(R@k)} score, and \textbf{Normalized Discounted accumulated Gain (N@k)}, measuring on top-$k$ recommendation results~\cite{wu2022disentangled,fan2023jointly}.  

\begin{table*}[h]
\centering
\vskip -0.1in
\caption{Fairness and recommendation utility of different methods on two backbone models over datasets with a binary-valued item attribute (evaluated on top-20 recommendation results). \textbf{RI} denotes the relative improvement of \mymodel{} over UNS. We highlight the best results in bold and the \underline{second best results} with underline.}
\label{tab: table_comparison22}
\vskip -0.10in
\scalebox{0.85}
{
\begin{tabular}{llllllllllllll}
\hline
\hline
                          &                  & \multicolumn{6}{c}{\textbf{ML1M-2}}                                                                                                                 & \multicolumn{6}{c}{\textbf{Amazon-2}}                                                                                   \\ \hline
Backbone                  & Method           & {\begin{tabular}[c]{@{}l@{}}Recall-\\ Disp.@20\end{tabular}}    & {\begin{tabular}[c]{@{}l@{}}Recall-\\ Min.@20\end{tabular}}     & \multicolumn{1}{l|}{\begin{tabular}[c]{@{}l@{}}Recall-\\ Avg.@20\end{tabular}}     & N@20            & P@20            & \multicolumn{1}{l|}{R@20}            & {\begin{tabular}[c]{@{}l@{}}Recall-\\ Disp.@20\end{tabular}}    & {\begin{tabular}[c]{@{}l@{}}Recall-\\ Min.@20\end{tabular}}     & \multicolumn{1}{l|}{{\begin{tabular}[c]{@{}l@{}}Recall-\\ Avg.@20\end{tabular}}}           & N@20            & P@20            & R@20            \\ \hline
\multirow{9}{*}{MF}       & UNS & 0.3179          & 0.2346          &      \multicolumn{1}{l|}{0.3386}          & 0.3487          & \underline{0.1611}          & \multicolumn{1}{l|}{0.4622}          & 0.6882          & 0.0386          & \multicolumn{1}{l|}{0.1238}       & 0.0780          & 0.0196          & 0.1471          \\
                          & NNCF             & 0.1689          & 0.2550          & \multicolumn{1}{l|}{0.3069}          & 0.3051          & 0.1386          & \multicolumn{1}{l|}{0.4157}          & \underline{0.5157}          & \underline{0.0594}          & \multicolumn{1}{l|}{0.1227}       & 0.0727          & 0.0187          & 0.1159                  \\
                          
                          & DNS             & 0.3065          & 0.2485          & \multicolumn{1}{l|}{\underline{0.3584}}          & \textbf{0.3765} & \textbf{0.1702} & \multicolumn{1}{l|}{\textbf{0.4974}} & 0.5937          & 0.0547          & \multicolumn{1}{l|}{\textbf{0.1347}}       & \textbf{0.0851} & \textbf{0.0209} & \textbf{0.1591}             \\
                          & FairStatic       & \underline{0.0958}          & \underline{0.3208}          & \multicolumn{1}{l|}{0.3548}          & 0.3409          & 0.1558          & \multicolumn{1}{l|}{0.4541}          & 0.5936          & 0.0501          & \multicolumn{1}{l|}{0.1232}       & 0.0769          & 0.0191          & 0.1426             \\ \cline{2-14} 
                          & Reg              & 0.1908          & 0.2808          & \multicolumn{1}{l|}{0.3471}          & 0.3156          & 0.1580          & \multicolumn{1}{l|}{0.4518}          & 0.5648          & 0.0534          & \multicolumn{1}{l|}{0.1227}       & 0.0751          & 0.0189          & 0.1409      \\
                          & DPR              & 0.0959          & 0.3147          & \multicolumn{1}{l|}{0.3481}          & 0.3134          & 0.1528          & \multicolumn{1}{l|}{0.4295}          & 0.5687          & 0.0480          & \multicolumn{1}{l|}{0.0963}       & 0.0610          & 0.0162          & 0.1283        \\
                          & FairGAN-1        & 0.3607          & 0.2056          & \multicolumn{1}{l|}{0.3217}          & 0.3267          & 0.1495          & \multicolumn{1}{l|}{0.4287}          & 0.6503          & 0.0467          & \multicolumn{1}{l|}{0.1285}       & 0.0672          & 0.0167          & 0.1219          \\ \cline{2-14}
                          & \textbf{FairNeg} & \textbf{0.0287} & \textbf{0.3659} & \multicolumn{1}{l|}{\textbf{0.3764}} & {\underline{0.3540}}    & 0.1609          & \multicolumn{1}{l|}{\underline{0.4734}}    & \textbf{0.5111} & \textbf{0.0638} & \multicolumn{1}{l|}{\underline{0.1305}} & \underline{0.0790}    & \underline{0.0198}    &\underline{0.1515}    \\ \cline{2-14}
                          & \textbf{RI}         & 90.97\%         & 55.97\%         & \multicolumn{1}{l|}{11.16\%}         & 1.53\%          & -0.12\%         & \multicolumn{1}{l|}{2.42\%}          & 25.73\%         & 65.28\%         & \multicolumn{1}{l|}{5.37\%}       & 1.26\%          & 1.02\%          & 2.99\%          \\ \hline
\multirow{9}{*}{LightGCN} & UNS              & 0.2085          & 0.2928          & \multicolumn{1}{l|}{0.3699}          & \underline{0.3745}          & \underline{0.1695}          & \multicolumn{1}{l|}{\underline{0.5029}}          & 0.6588          & 0.0470          & \multicolumn{1}{l|}{\underline{0.1378}}                & \underline{0.0874}          & \underline{0.0217}          & \underline{0.1670}          \\
                          & NNCF             & 0.1600          & 0.2979          & \multicolumn{1}{l|}{0.3547}          & 0.3576          & 0.1596          & \multicolumn{1}{l|}{0.4785}          & \underline{0.5473}          & \underline{0.0594}          & \multicolumn{1}{l|}{0.1313}                & 0.0825          & 0.0201          & 0.1556          \\
                          & DNS              & 0.2567          & 0.2728          & \multicolumn{1}{l|}{0.3670}          & \textbf{0.3794} & \textbf{0.1712} & \multicolumn{1}{l|}{\textbf{0.5054}} & 0.6704          & 0.0453          & \multicolumn{1}{l|}{0.1375}                & \textbf{0.0901} & \textbf{0.0220} & \textbf{0.1676} \\
                          & FairStatic       & \underline{0.0760}          & 0.3505          & \multicolumn{1}{l|}{\underline{0.3853}}          & 0.3495          & 0.1570          & \multicolumn{1}{l|}{0.4753}          & 0.5686          & 0.0591          & \multicolumn{1}{l|}{0.1370}                & 0.0847          & 0.0211          & 0.1619          \\ \cline{2-14} 
                          & Reg              & 0.1446          & 0.3203          & \multicolumn{1}{l|}{0.3744}          & 0.3677          & 0.1675          & \multicolumn{1}{l|}{0.4998}          & 0.5989          & 0.0544          & \multicolumn{1}{l|}{0.1356}                & 0.0849          & 0.0210          & 0.1599          \\
                          & DPR              & 0.0768          & \underline{0.3514}          & \multicolumn{1}{l|}{0.3806}          & 0.3614          & 0.1659          & \multicolumn{1}{l|}{0.4938}          & 0.5900          & 0.0551          & \multicolumn{1}{l|}{0.1343}                & 0.0842          & 0.0208          & 0.1584          \\
                          & FairGAN-2        & 0.4117          & 0.1941          & \multicolumn{1}{l|}{0.3300}          & 0.3501          & 0.1561          & \multicolumn{1}{l|}{0.4633}          & 0.6620          & 0.0457          & \multicolumn{1}{l|}{0.1351}                & 0.0677          & 0.0169          & 0.1229          \\ \cline{2-14} 
                          & \textbf{FairNeg} & \textbf{0.0682} & \textbf{0.3587} & \multicolumn{1}{l|}{\textbf{0.3865}} & {0.3721}    & 0.1675          & \multicolumn{1}{l|}{{0.5020}}    & \textbf{0.5217} & \textbf{0.0661} & \multicolumn{1}{l|}{{\textbf{0.1383}}} & {0.0860}    & {0.0211}    & {0.1598}    \\ \cline{2-14}
                          & \textbf{RI}      & 67.29\%         & 22.51\%         & \multicolumn{1}{l|}{4.47\%}          & -0.64\%         & -1.18\%         & \multicolumn{1}{l|}{-0.18\%}         & 20.81\%         & 40.64\%         & \multicolumn{1}{l|}{0.36\%}                & -1.64\%         & -2.76\%         & -4.31\%         \\ \hline
\end{tabular}
}
\vskip -0.10in
\end{table*}

\begin{table*}[h]
\centering
\caption{Fairness and recommendation utility of different methods on two backbone models over datasets with a multi-valued item attribute (evaluated on top-20 recommendation results). \textbf{RI} denotes the relative improvement of \mymodel{} over UNS. We highlight the best results in bold and the \underline{second best results} with underline.}
\label{tab: table_comparison33}
\vskip -0.10in
\scalebox{0.85}
{
\begin{tabular}{llllllllllllll}
\hline
\hline
                          &                  & \multicolumn{6}{c}{\textbf{ML1M-4}}                                                                                                                 & \multicolumn{6}{c}{\textbf{Amazon-4}}                                                                                                \\ \hline
Backbone                  & Method           & {\begin{tabular}[c]{@{}l@{}}Recall-\\ Disp.@20\end{tabular}}    & {\begin{tabular}[c]{@{}l@{}}Recall-\\ Min.@20\end{tabular}}     & \multicolumn{1}{l|}{\begin{tabular}[c]{@{}l@{}}Recall-\\ Avg.@20\end{tabular}}     & N@20            & P@20            & \multicolumn{1}{l|}{R@20}            & {\begin{tabular}[c]{@{}l@{}}Recall-\\ Disp.@20\end{tabular}}    & {\begin{tabular}[c]{@{}l@{}}Recall-\\ Min.@20\end{tabular}}     & \multicolumn{1}{l|}{{\begin{tabular}[c]{@{}l@{}}Recall-\\ Avg.@20\end{tabular}}}           & N@20            & P@20            & R@20            \\ \hline
\multirow{9}{*}{MF}       & UNS &0.2408           & 0.1963          &      \multicolumn{1}{l|}{0.2931}          & 0.2918          & 0.1474          & \multicolumn{1}{l|}{0.3765}          & 0.6513          & 0.0256          & \multicolumn{1}{l|}{0.0921}       & 0.0737          & 0.0318           & 0.1020           \\

                          & NNCF             & 0.1603                              & 0.2336                              & \multicolumn{1}{l|}{\underline{0.3008}}        & 0.3006                           & 0.1485                      & \multicolumn{1}{l|}{0.3884}          & \multicolumn{1}{l}{\underline{0.5240}}  & \multicolumn{1}{l}{0.0335} & \multicolumn{1}{l|}{0.0803}              & \multicolumn{1}{l}{0.0599} & \multicolumn{1}{l}{0.0270}  & \multicolumn{1}{l}{0.0854} \\

                          & DNS              & 0.2350                               & 0.2256                              & \multicolumn{1}{l|}{0.3146}        & \textbf{0.3203}                           & \textbf{0.1573}                      & \multicolumn{1}{l|}{\textbf{0.4148}}          & \multicolumn{1}{l}{0.6596} & \multicolumn{1}{l}{0.0243} & \multicolumn{1}{l|}{0.0948}              & \multicolumn{1}{l}{\underline{0.0751}} & \multicolumn{1}{l}{\underline{0.0322}} & \multicolumn{1}{l}{\underline{0.1031}} \\

                        & FairStatic       & 0.1154                              & \underline{0.2615}                              & \multicolumn{1}{l|}{0.2931}        & 0.2722                           & 0.1394                      & \multicolumn{1}{l|}{0.3624}          & \multicolumn{1}{l}{0.5389} & \multicolumn{1}{l}{\underline{0.0366}} & \multicolumn{1}{l|}{0.0979}               & \multicolumn{1}{l}{0.0678} & \multicolumn{1}{l}{0.0302} & \multicolumn{1}{l}{0.0953} \\ \cline{2-14} 

                          & DPR          & \underline{0.0756}                              & 0.2550                               & \multicolumn{1}{l|}{0.2878}         & 0.2622                           & 0.1372                      & \multicolumn{1}{l|}{0.3549}          & \multicolumn{1}{l}{0.6257} & \multicolumn{1}{l}{0.0232} & \multicolumn{1}{l|}{0.0725}              & \multicolumn{1}{l}{0.0514} & \multicolumn{1}{l}{0.0238} & \multicolumn{1}{l}{0.0708} \\

                          & FairGAN-1        & 0.3012                              & 0.1902                              & \multicolumn{1}{l|}{0.2945}        & 0.292                            & 0.1433                      & \multicolumn{1}{l|}{0.3762}          & \multicolumn{1}{l}{0.6565} & \multicolumn{1}{l}{0.0254} & \multicolumn{1}{l|}{\underline{0.0990}}                 & \multicolumn{1}{l}{\textbf{0.0768}} & \multicolumn{1}{l}{\textbf{0.0324}} & \multicolumn{1}{l}{0.1013} \\ \cline{2-14} 
                          
                          & \textbf{FairNeg} & \textbf{0.0318} & \textbf{0.2998} & \multicolumn{1}{l|}{\textbf{0.3134}} & {\underline{0.3024}}    &\underline{0.1508}           & \multicolumn{1}{l|}{\underline{0.3925}}    & \textbf{0.5061} & \textbf{0.0439} & \multicolumn{1}{l|}{\textbf{0.0999}} & {0.0724}    & {0.0321}    &{\textbf{0.1035}}    \\ \cline{2-14}
                          
                          & \textbf{RI}         & 86.78\%         & 52.73\%         & \multicolumn{1}{l|}{6.90\%}         & 3.63\%          & 2.31\%         & \multicolumn{1}{l|}{4.25\%}          & 22.29\%         & 71.48\%         & \multicolumn{1}{l|}{8.47\%}       & -1.76\%          & 0.94\%          & 1.47\%          \\ \hline
\multirow{9}{*}{LightGCN} & UNS              & 0.1794          & 0.2455          & \multicolumn{1}{l|}{0.3236}          & \underline{0.3294}          & \underline{0.1603}          & \multicolumn{1}{l|}{\underline{0.4279}}          & 0.6799          & 0.0247          & \multicolumn{1}{l|}{0.1015}          & \underline{0.0801}          & \underline{0.0344}          & \underline{0.1111}          \\

                          & NNCF              & 0.1091           &  0.2571         & \multicolumn{1}{l|}{0.3060}          & 0.3095  & 0.1498  & \multicolumn{1}{l|}{0.4033}          & \underline{0.5564}  & 0.0342  &  \multicolumn{1}{l|}{0.1015}          & 0.0754 & 0.0327 & 0.1044 \\
                         
                          & DNS              & 0.1806          & 0.2463           & \multicolumn{1}{l|}{\underline{0.3259}}          & \textbf{0.3352} & \textbf{0.1615} & \multicolumn{1}{l|}{\textbf{0.4374}} &  0.6711         & 0.0258         & \multicolumn{1}{l|}{\textbf{0.1053}}          & \textbf{0.0822} & \textbf{0.0353} & \textbf{0.1125} \\
                          
                          & FairStatic       & 0.1022          & 0.2819           & \multicolumn{1}{l|}{0.3208}          & 0.3156          & 0.1522           & \multicolumn{1}{l|}{0.4131}          & 0.5679          & \underline{0.0410}           & \multicolumn{1}{l|}{0.1024}          & 0.0770          & 0.0338           & 0.1062           \\ \cline{2-14} 
                          
                          & DPR              & \underline{0.0495}         & \underline{0.2994}          & \multicolumn{1}{l|}{0.3207}          & 0.3128         & 0.1552          & \multicolumn{1}{l|}{0.4084}          & 0.5997         & 0.0289          & \multicolumn{1}{l|}{0.1017}          & 0.0773         & 0.0339          & 0.1075           \\
                          & FairGAN-2        & 0.3251         & 0.1799          & \multicolumn{1}{l|}{0.2999}          & 0.3064           & 0.1447          & \multicolumn{1}{l|}{0.3980}          & 0.5768          & 0.0324          & \multicolumn{1}{l|}{0.1003}          & 0.0742          & 0.0320         & 0.1004           \\ \cline{2-14}
                          
                          & \textbf{FairNeg} & \textbf{0.0394} & \textbf{0.3117} & \multicolumn{1}{l|}{\textbf{0.3267}} & {0.3232}    & 0.1565          & \multicolumn{1}{l|}{{0.4222}}    & \textbf{0.5062} & \textbf{0.0420} & \multicolumn{1}{l|}{{\underline{0.1042}}} & {0.0792}    & {0.0340}    & {0.1097}    \\ \cline{2-14} 
                          & \textbf{RI}      & 78.04\%         & 26.97\%         & \multicolumn{1}{l|}{97.35\%}          & -1.88\%         & -2.37\%         & \multicolumn{1}{l|}{-1.33\%}         & 25.55\%         & 70.04\%         & \multicolumn{1}{l|}{2.66\%}                & -1.17\%         & -1.25\%         & -1.26\%         \\ \hline
\end{tabular}
}
\vskip -0.05in
\end{table*}

\subsubsection{Parameter Settings} 
Due to the limited space, more implementation details are presented in the Appendix Section~\ref{sec:params}.

\subsubsection{Baselines} 
We compare \mymodel{} with four representative negative sampling methods and three state-of-the-art item-side fairness-aware recommendation methods. The method descriptions are as follows: 

\begin{itemize}[leftmargin=*] 
\item  \textbf{UNS}~\cite{rendle2012bpr}: a widely used negative sampling approach using a static uniform distribution in the sampling process.
\item \textbf{NNCF}~\cite{rendle2014improving}: a static negative sampling approach using a item-popularity-based distribution.
\item  \textbf{DNS}~\cite{zhang2013dns}: a dynamic negative sampling method that assigns higher probability to samples with higher ranking order.
\item  \textbf{FairStatic}: a simplified variant of \mymodel{} based on a static sampling distribution, as described in the introduction.
\item  \textbf{Reg}~\cite{kamishima2017reg}: a typical debiasing method for datasets with a binary-valued attribute, which penalizes the squared difference between the average scores of two groups for all positive user-item pairs.

\item  \textbf{DPR}~\cite{zhu2020measuring}: an advanced item group fairness method, which uses score-based adversarial-learning to decouple the predicted user-item preference score with the item's attribute.

\item  \textbf{FairGAN}~\cite{li2022fairgan}: a GAN-based algorithm that fairly allocates exposure to each item with high user utility preservation.
\end{itemize}

\subsection{Performance Comparison of Recommender Systems}

The comparison experiments are conducted on two backbone models (\textbf{MF} and \textbf{LightGCN}) separately. Each experiment is conducted five times with different random seeds, and we report the average result. Based on top-20 recommendation results, the item group fairness and their recommendation performance on four datasets are respectively shown in Table~\ref{tab: table_comparison22},~\ref{tab: table_comparison33}. The Reg method can only work for binary group cases, so we only report its performance on \textbf{ML1M-2} and \textbf{Amazon-2} datasets. FairGAN has a fairness strictness hyperparameter. Thus we set two different values on the two backbone models separately so that the utility is close to other methods, denoted as FairGAN-1 and FairGAN-2.
We have the following observations from the results (the evaluation of top-30 recommendation results are presented in the Appendix Section~\ref{sec:comparison}, which also aligns with the observations):

First, compared with \mymodel{}, most methods provide biased recommendation results regarding the Recall-Disp, Recall-min, and Recall-Avg metrics. 
Second, compared with the methods that do not explicitly consider group fairness (e.g., NNCF and DNS), fairness-aware methods usually yield better group-wise recall parity. 
Among them, the adversarial learning-based method usually performs better than others. 
It is worth noting that FairStatic shows competitive fairness results, which indicates that optimizing negative sampling distribution with debias considerations also brings substantial fairness improvements. 

Third, \mymodel{} consistently outperforms other methods on different backbone models over four datasets regarding its effectiveness in mitigating group-wise recall performance unfairness. 
At last, \mymodel{}  can achieve superior fairness while preserving good recommendation utility on the user side. Even though DNS provides the best recommendation utility, it ignores the huge performance disparity among item groups. 
To conclude, these results validate that our approach \mymodel{} can effectively improve fairness in group-wise recall performance with minor recommendation utility loss.

\subsection{Ablation Study}

\begin{figure*}[h]
\centering
\vskip -0.05in
\subfigure[MF:Fairness on ML1M-4]{\label{fig:str}\includegraphics[width=0.24\linewidth]{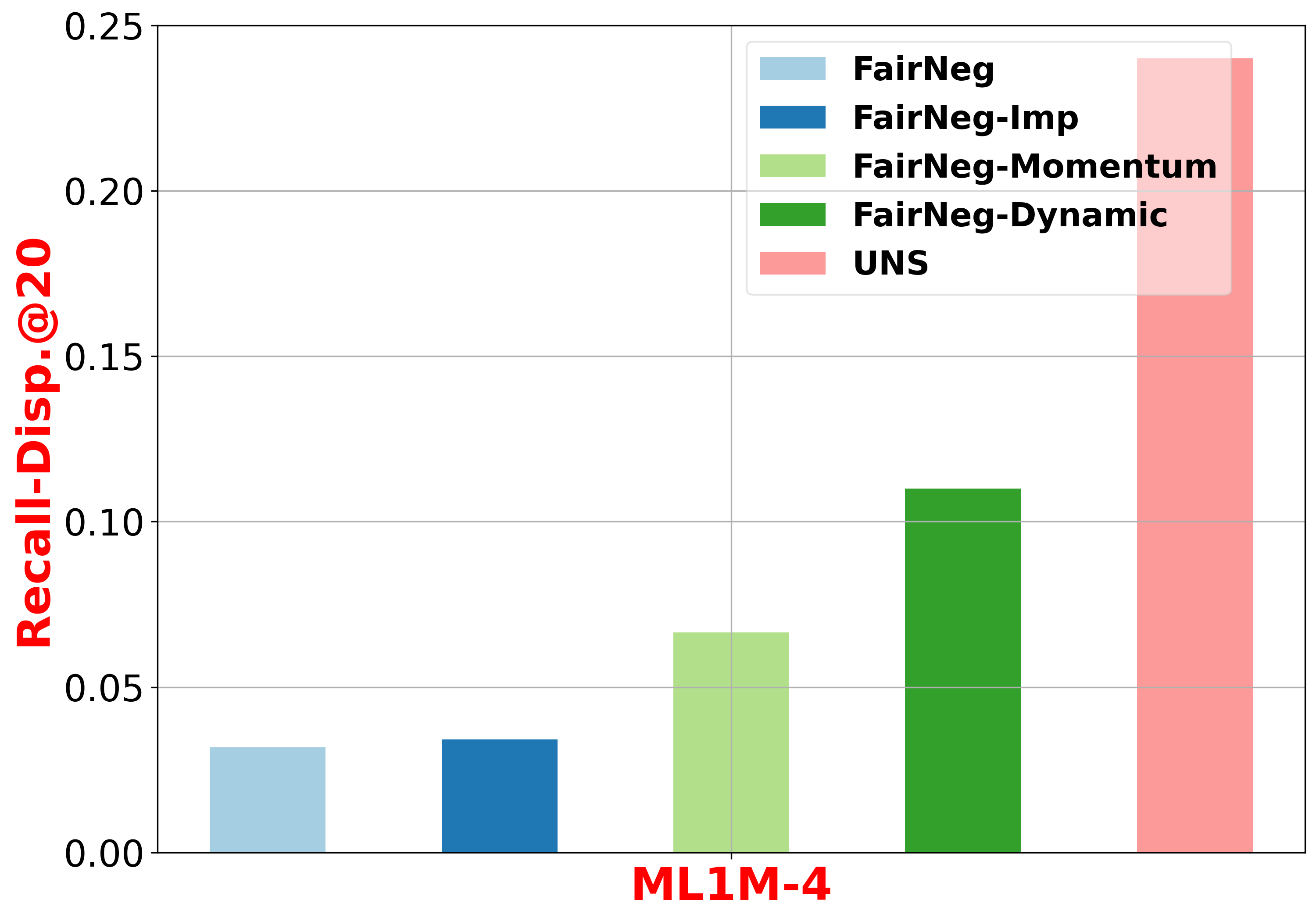}}
\subfigure[MF:Utility on ML1M-4]{\label{fig:fs}\includegraphics[width=0.24\linewidth]{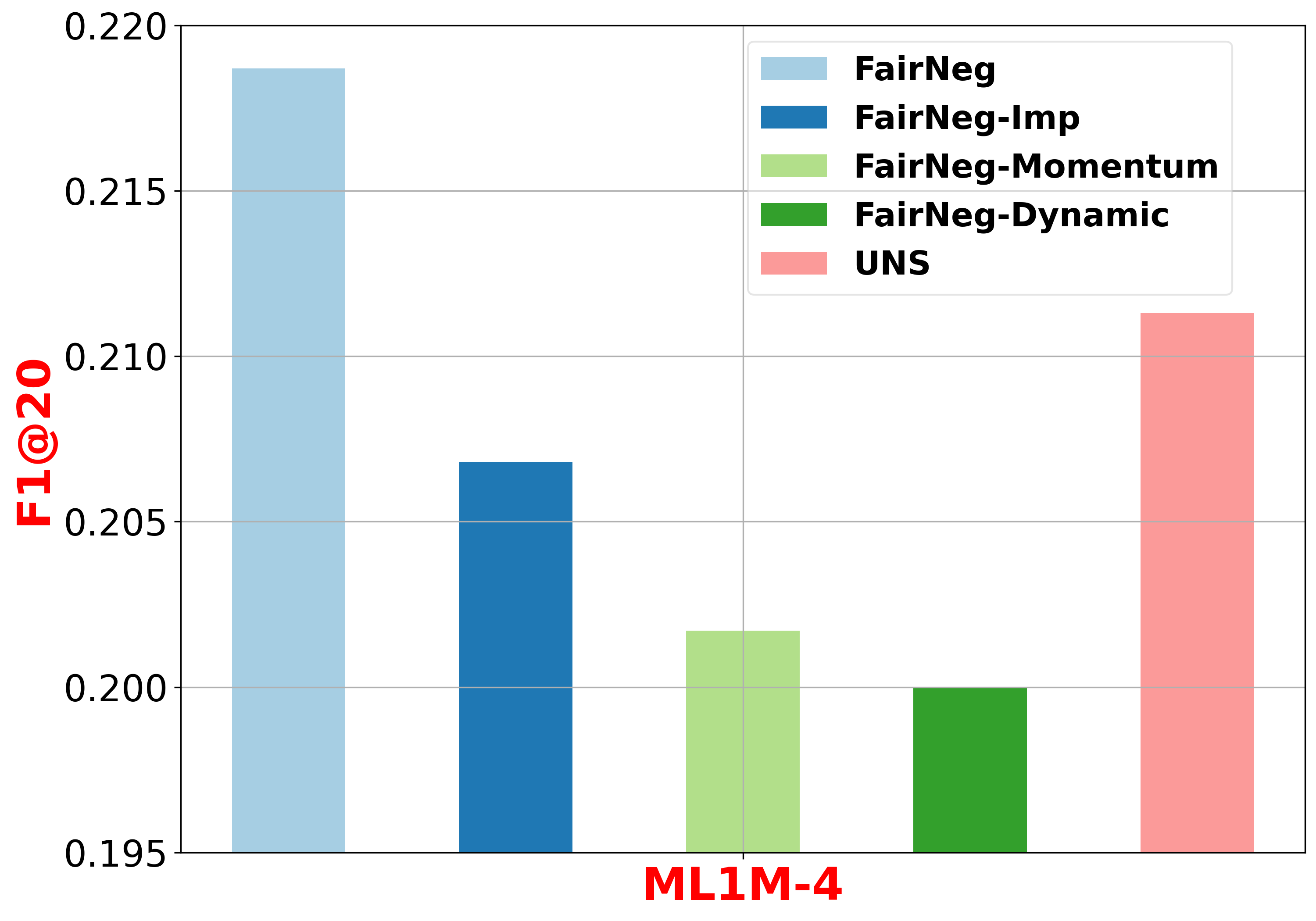}}
\subfigure[LightGCN:Fairness on ML1M-4]{\label{fig:str}\includegraphics[width=0.24\linewidth]{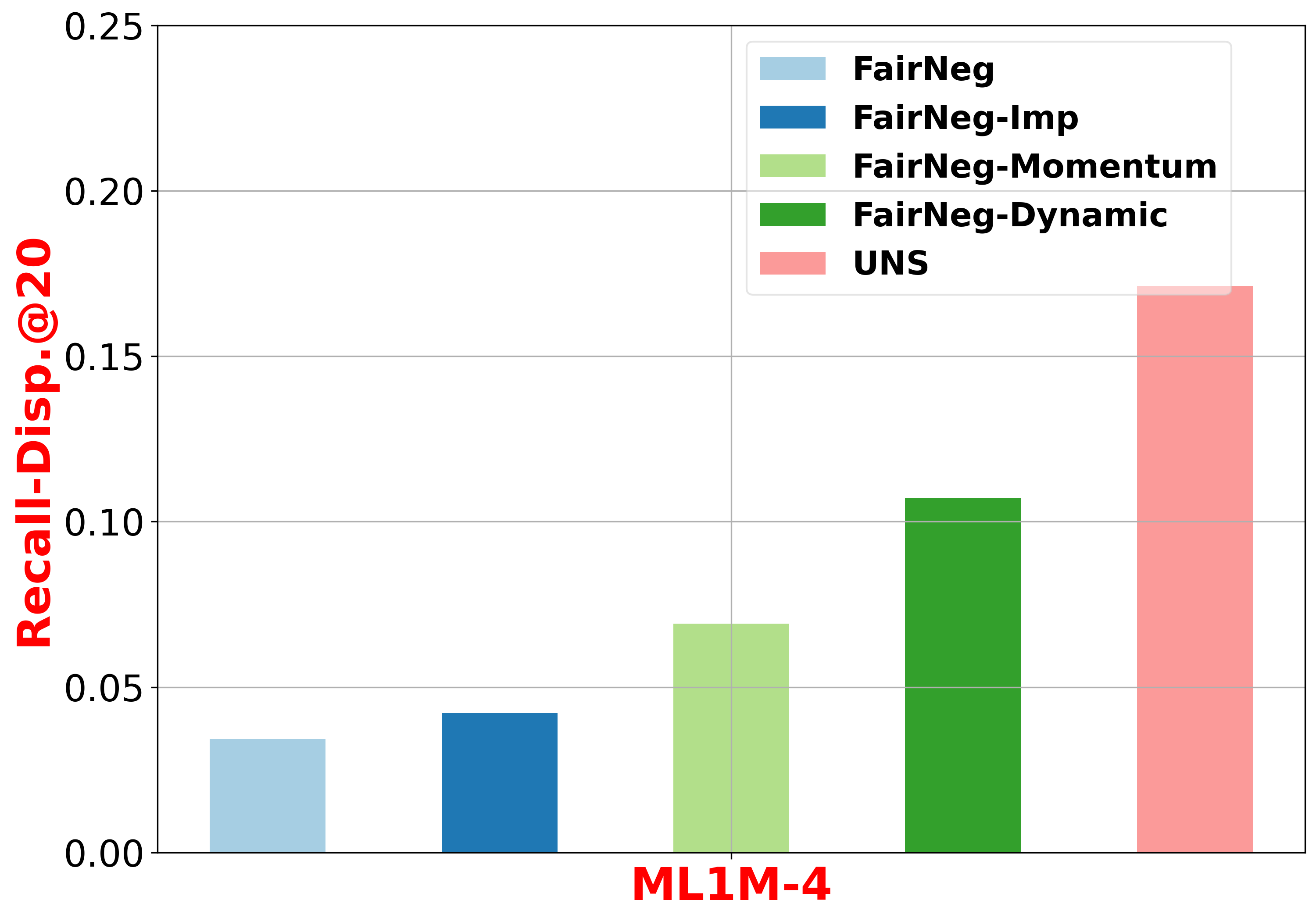}}
\subfigure[LightGCN:Utility on ML1M-4]{\label{fig:fs}\includegraphics[width=0.24\linewidth]{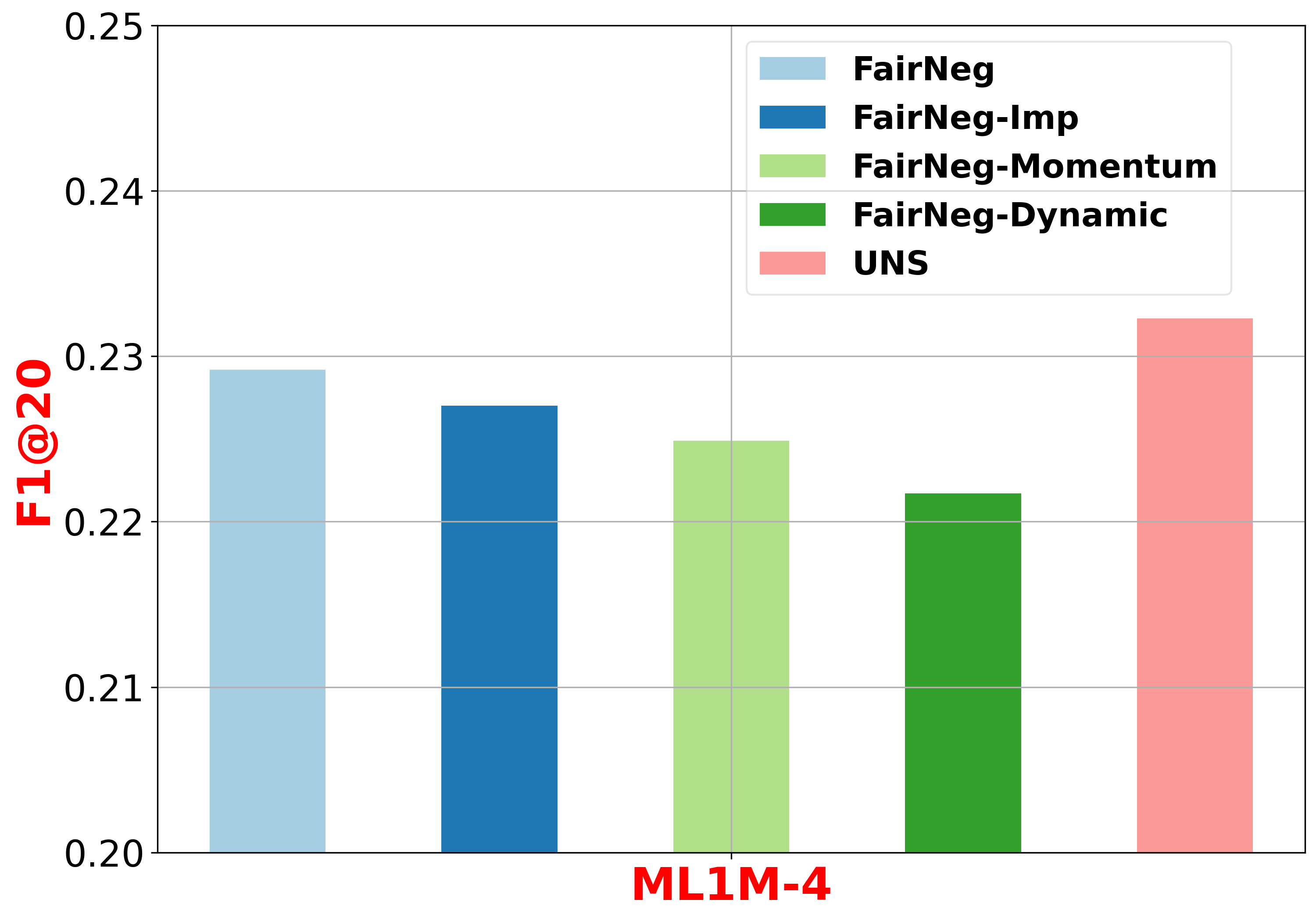}}
\vskip -0.150in
\caption{ The F1@20 and Recall-Disp@20 of \mymodel{} and its variants on two backbones over the ML1M-4 dataset.}
\label{fig:ab_fig1}
 \vskip -0.05in
\end{figure*}

Next, we conduct ablation studies to verify the effectiveness of each component in \mymodel{}. We compare \mymodel{} with its variants, where one component are removed (i.e., \emph{Dynamic fair negative sampling} (\mymodel{}-Dynamic), \emph{Importance-aware sampling} (\mymodel{}-Imp), and \emph{adaptive momentum update of group NS probability} (\mymodel{}-Momentum)). The uniform negative sampling (UNS) method served as the baseline. All the methods are evaluated on both MF and LightGCN backbone. 

For brevity, we report each variant's fairness performance (Recall-Disp@20) and the recommendation performance (F1@20) over the ML1M-4 dataset, as presented in Figure~\ref{fig:ab_fig1} (results on Amazon-4 dataset are presented in the Appendix Section~\ref{sec:ablation study}). 

We have the following findings from the ablation studies. First, the dynamic sampling mechanism (which can be viewed as FairStatic) is vital for fairness improvement and utility-preserving, indicating the importance of adjusting the sampling distribution adaptively. 
Second, incorporating the importance-aware negative sampling probability distribution is vital for utility preservation, since the informativeness of negative items can be enhanced with such a strategy, further contributing to feature representation learning. 
Third, removing the momentum mechanism when updating the group-level sampling probability dampens fairness performance, which indicates that the fairness optimization process is unstable and the gradient noise needs to be considered. 
Fourth, \mymodel{} achieves the best utility-fairness tradeoff on two backbones, proving that our proposed method's effectiveness is not limited to a specific model architecture.

\section{Related Work}
In this section, we briefly review related work, including fair recommendations and negative sampling.

\noindent \textbf{Fair Recommendations}.
Fair recommendations have drawn tremendous attention in recent years. The fairness notions can be defined from the user side~\cite{li2021user, leonhardt2018user,fan2022comprehensive} or the item side~\cite{ge2021towards}, or both~\cite{burke2017multisided}, which makes fair recommendation quite challenging. On the item side, many existing works focus on exposure fairness in the ranking results~\cite{singh2018fairness, patro2020fairrec, morik2020controlling}, for instance, in~\cite{morik2020controlling}, a configurable allocation-of-exposure scheme is proposed to overcome richer-get-richer dynamics in the ranking system.
However, these methods only study exposure opportunities allocation of items, and the true user preferences are not considered. 
When items in certain groups have lower recommendation probabilities, given that the items are liked by users, the recommendation model suffers from item under-recommendation bias. A series of works have been proposed to achieve performance (i.e., recall) parity among different item groups. These methods can be divided into two types: (i) fairness-constrained regularization method~\cite{kamishima2017reg}, where the average score difference of item groups is formed as a regularizer to penalize the group-wise performance disparity; (ii) adversary learning methods~\cite{liu2022dual,zhu2020measuring,zhang2022fairness}, where a discriminator classifies the group attribute based on the predicted user-item relevance scores and the recommendation model prevents the discriminator from correctly classifying the groups. 
Unlike these methods, our approach proposes a fairly adaptive negative sampling technique, which addresses group unfairness from negative sampling perspective.

\noindent \textbf{Negative Sampling}.
Negative sampling techniques~\cite{mikolov2013distributed} are frequently used in pairwise algorithms~\cite{yu2018multiple}, and we can group them into static negative sampling and hard negative sampling. The former (i.e., uniform sampling) utilizes a fixed sampling distribution and provides high efficiency. The latter~\cite{zhang2013dns, wang2017irgan, ding2020simplify} selects the negative instances that are similar to positive instances in the embedding space in a dynamic way. For example, in~\cite{park2019adversarial}, they adopt adversarial learning to mine hard examples, which brings more information to model training. In~\cite{ding2020simplify}, they leverage a high-variance based method to choose high-quality negative instances in a reliable way. However, these methods are mainly designed to maximize the overall utility~\cite{wan2022cross}, while neglecting the data bias (group-wise) on the item side. Our work makes the first attempt to build the connection between item group fairness optimization and negative sampling.
\section{Conclusion}
Item-oriented group performance fairness is an essential factor for building trustworthy recommender systems~\cite{fan2022comprehensive,liu2021trustworthy}. In this work, we propose a novel \textbf{Fair}ly adaptive \textbf{Neg}ative sampling framework (FairNeg) to alleviate the adverse impact of negative sampling on training recommendation models. 
Based on the pairwise training paradigm, we introduce the fairness perception module to measure the recall performance disparity and then adjust the group-wise sampling probability with an adaptive momentum mechanism. Furthermore, we introduce the mixup mechanism for combining fairness-aware and importance-related sampling distribution, which aims to jointly consider representation learning and group fairness optimization. Extensive experiments show that our method outperforms state-of-the-art debiasing approaches regarding fairness performance by significant margins and yields better fairness-performance tradeoffs. 
In the future, we plan to investigate the possibility of leveraging the framework to be compatible with more group fairness metrics in recommendations.
\section*{Acknowledgment}
The research described in this paper has been partly supported by NSFC (Project No. 62102335),  a General Research Fund from the Hong Kong Research Grants Council (Project No. PolyU 15200021 and PolyU 15207322), and internal research funds from The Hong Kong Polytechnic University (Project No. P0036200, P0042693, and P0043302). This research was also supported by the InnoHK project. Dr. Zitao Liu was partly supported by  Key Laboratory of Smart Education of Guangdong Higher Education Institutes, Jinan University (2022LSYS003).

\balance
\bibliographystyle{ACM-Reference-Format}
\bibliography{acm}

\newpage
\balance
\appendix
\section{Appendix}
In this appendix, we provide the necessary information for reproducing our insights and experimental results. 

\subsection{Group-wise Statistics of Datasets}\label{sec:statistics}
The statistics of the four datasets
(i.e., feedback and item number) are illustrated in Table~\ref{tab:datasets_appendix}. We can observe that the item number distribution is fairly imbalanced on these datasets. Compared with ML1M dataset, Amazon dataset has lower density since the average positive feedback of each item group is much lower.

\begin{table}[h]
  \centering
   \vskip -0.1in
  \caption{Group statistics in the four datasets.}
 \vskip -0.1in
  \resizebox{0.8\columnwidth}{!}{
    \begin{tabular}{l|l|lll}
    \hline
                          & Group      & \#Item & \#Feedback & $\frac{\#Feedback}{\#Item}$ \\ \hline
    \multirow{2}{*}{ML1M-2}   & Sci-Fi     & 211    & 136,816     & 648             \\
                          & Horror     & 261    & 57,794      & 221             \\ \hline
    \multirow{2}{*}{Amazon-2} & Toy        & 1,307   & 30,695      & 23              \\
                          & Grocery    & 1,180   & 53,961      & 45              \\ \hline
    \multirow{4}{*}{ML1M-4}         & Sci-Fi     & 140    & 80,200      & 573             \\
                          & Adventure  & 129    & 58,679      & 455             \\
                          & Childrens' & 162    & 48,974      & 302             \\
                          & Horror     & 254    & 54,566      & 215             \\ \hline
    \multirow{4}{*}{Amazon-4}         & Grocery    & 708    & 49,095      & 69              \\
                          & Office     & 864    & 37,580      & 43              \\
                          & Pet        & 437    & 14,693      & 34              \\
                          & Tool       & 1,052   & 28,065      & 27              \\ \hline
    \end{tabular}%
    }
  \label{tab:datasets_appendix}%
 \vskip -0.10in
\end{table}%

\begin{figure}[b]
\vskip -0.15in
\centering
\subfigure[MF:Fairness on AMAZON-4]{\label{fig:fs}\includegraphics[width=0.48\columnwidth]{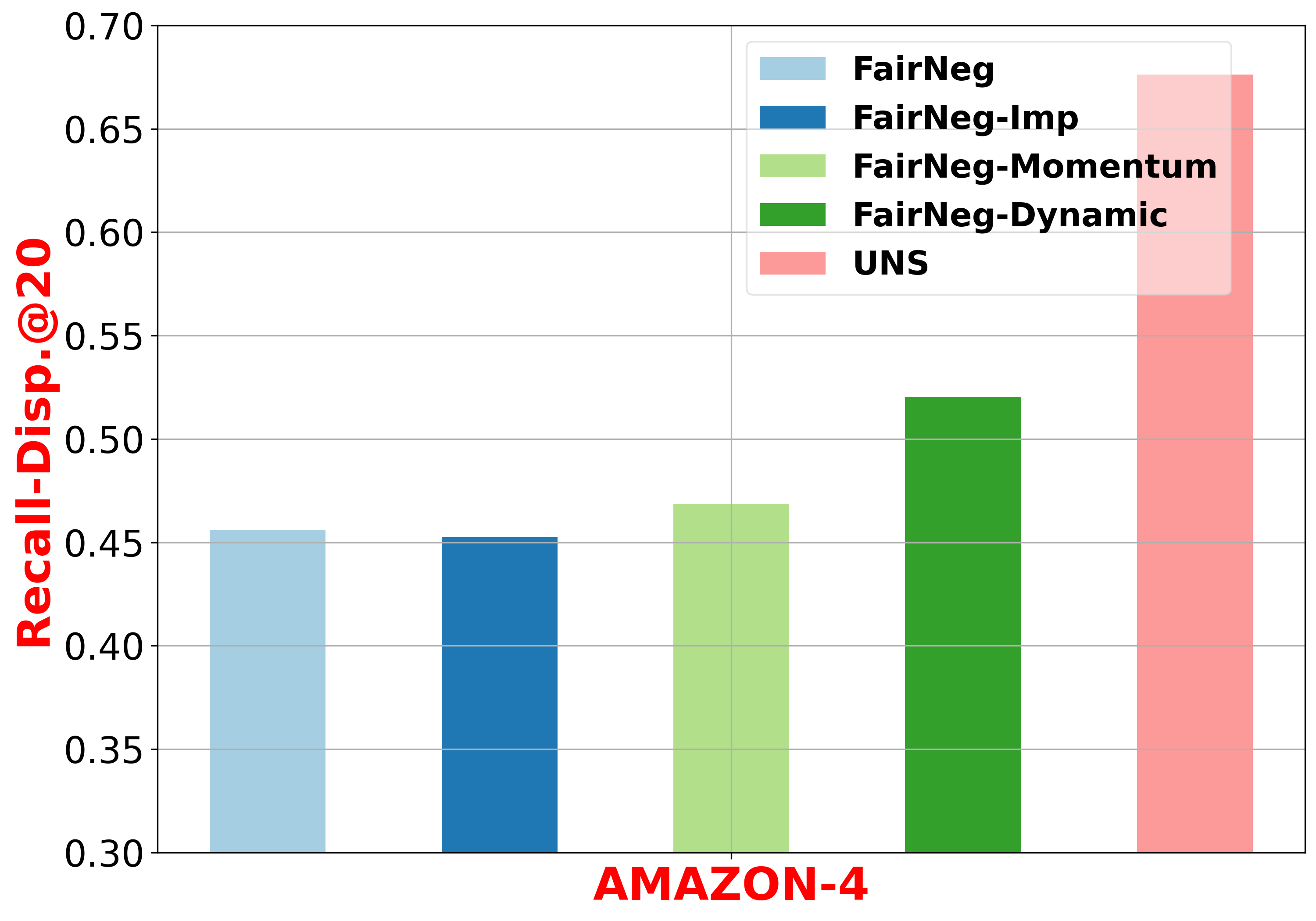}}
\subfigure[MF:Utility on AMAZON-4]{\label{fig:fs}\includegraphics[width=0.48\columnwidth]{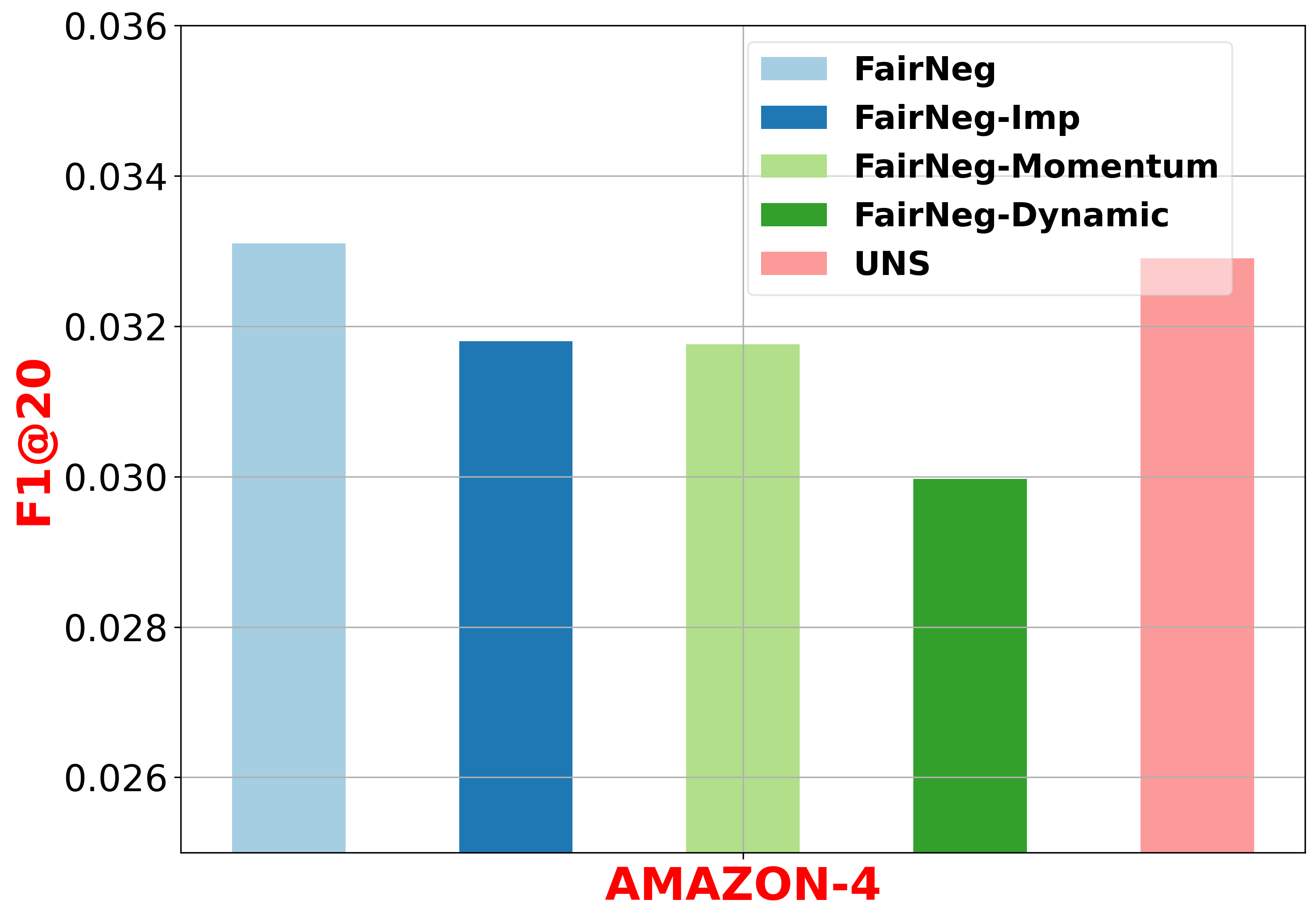}}
\subfigure[LightGCN:Fairness on AMAZON-4]{\label{fig:fs}\includegraphics[width=0.48\columnwidth]{figs/experiments/multiple_yaxis_ML1M4_LGN_Recall-Disp1.png}}
\subfigure[LightGCN:Utility on AMAZON-4]{\label{fig:fs}\includegraphics[width=0.48\columnwidth]{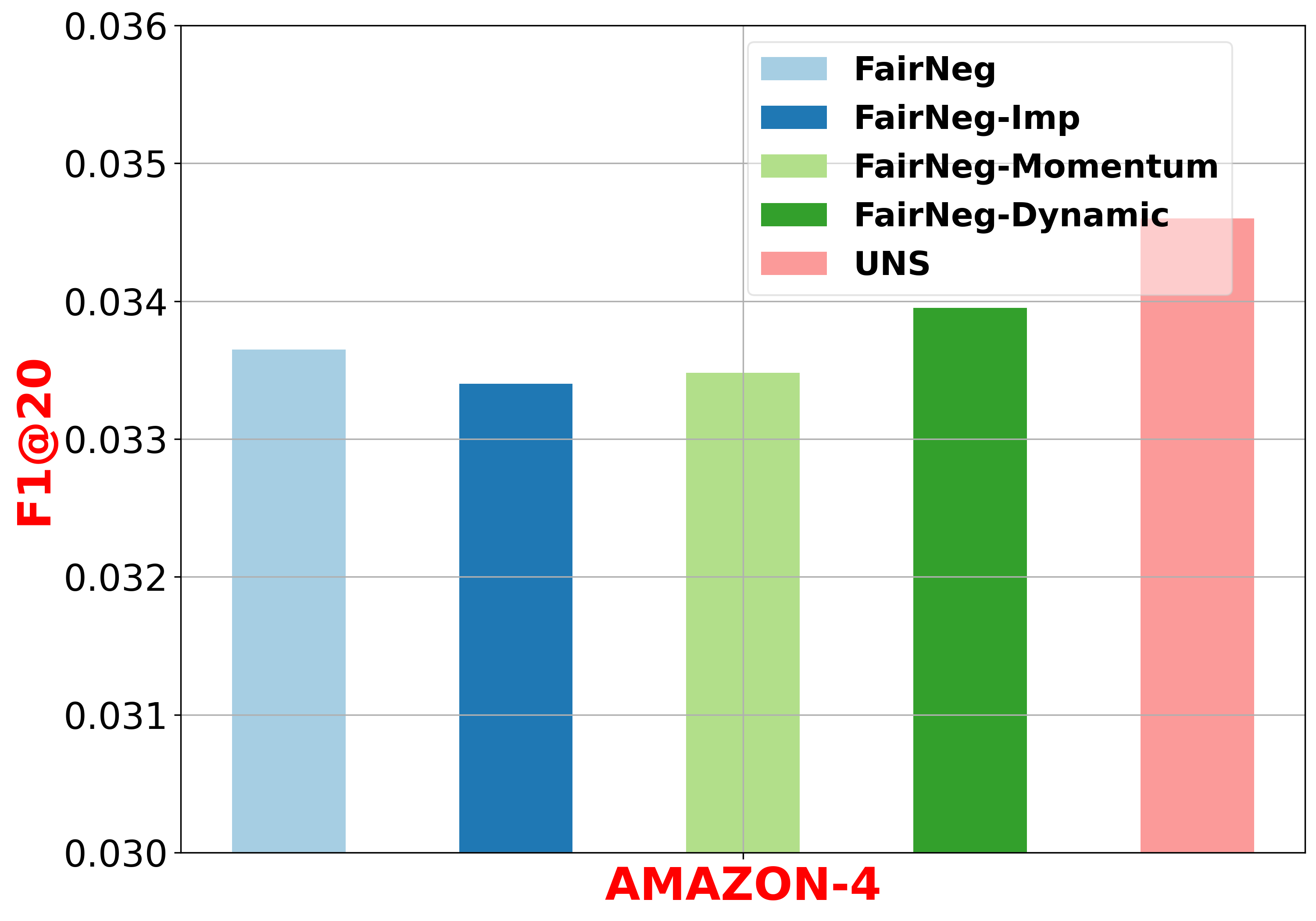}}
\vskip -0.15in
\caption{The Fairness/Recommendation utility comparison between \mymodel{} and its variants over amazon-4 dataset (using MF/LightGCN as backbone). }
\label{fig:ab_fig2}
\end{figure}

\subsection{{Implementation Details}}\label{sec:params}
All the experiments are based on two representative recommender models: MF~\cite{rendle2012bpr} and LightGCN~\cite{he2020lightgcn} optimized by Bayesian Personalized Ranking (BPR) loss, which do not explicitly take fairness criterion into considerations in their training paradigm. 
    Our proposed method \mymodel{} is implemented on the basis of PyTorch and Adam optimization. The baseline models for comparison are implemented on Tensorflow with Adam optimizer. We tune the hyper-parameters involved in the models with the validation set: the hidden dimensions are searched over [10, 80]  and the weight of $L_{2}$ regularizer is searched over \{0.01, 0.05, 0.1, 0.5\}. 
    Hyper-parameters for our proposed methods include: the momentum coefficient $\gamma$ is searched over [0, 0.2] with step 0.05 and the utility-fairness trade-off parameter $\beta$ is searched over [0.1, 0.9] with step 0.1. The learning rate $\alpha$ in the outer optimizer is set as 0.1, temperature $\tau$ in the hard-ware probability calculation is set as 0.4.
    Hyper-parameters defaults for the architecture of recommender system include: the negative sampling ratio set as 1 and mini-batch size set as 1024 for all methods in all experiments. The learning rate for methods using MF as backbone is 0.01 and 0.001 for those using LightGCN. The best models are selected based on the performance on the validation set within 100 epochs.

\begin{figure}[h]
\centering
 \vskip -0.1in
\subfigure[LightGCN:Fairness on ML1M-4 ]{\label{fig:str}\includegraphics[width=0.42\linewidth]{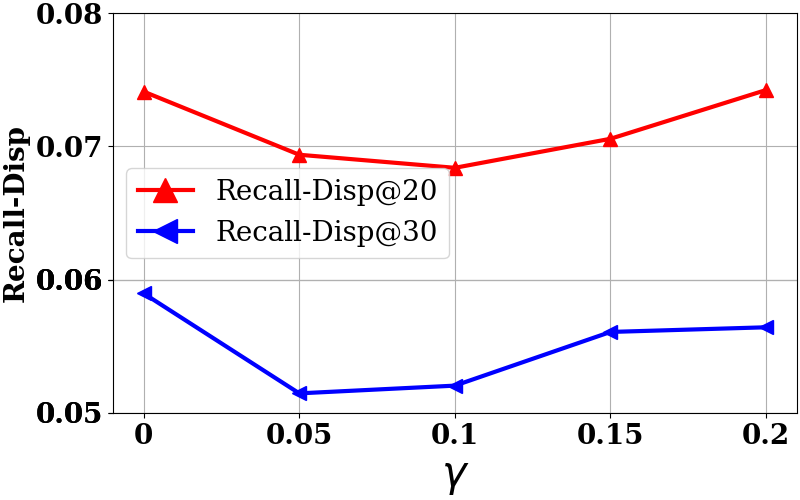}}
\subfigure[LightGCN:Utility on ML1M-4 ]{\label{fig:fs}\includegraphics[width=0.4\linewidth]{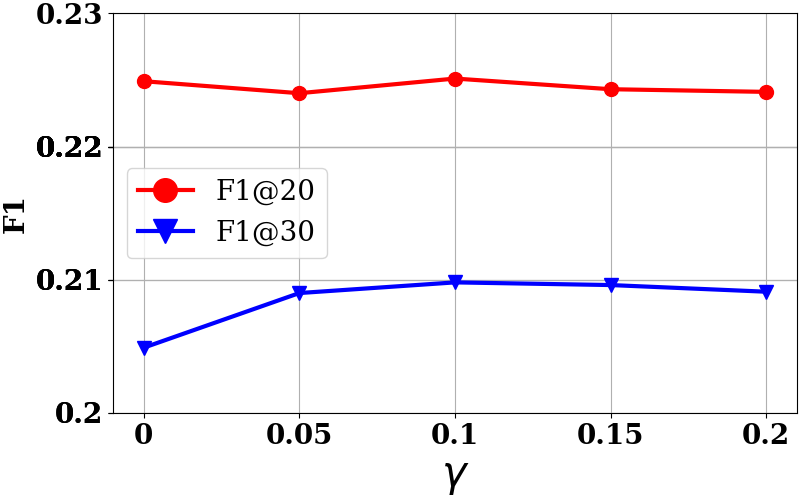}}
\vskip -0.15in
\caption{The fairness and utility w.r.t different $\gamma$.}
\label{fig:hyperparam_fig1}
\end{figure}

\begin{figure}[h]
\centering
 \vskip -0.10in
\subfigure[LightGCN:Fairness on ML1M-4 ]{\label{fig:str}\includegraphics[width=0.42\linewidth]{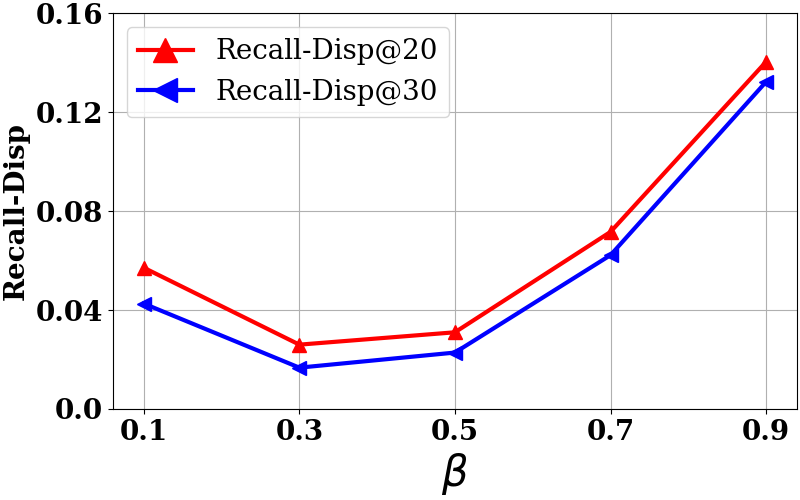}}
\subfigure[LightGCN:Utility on ML1M-4]{\label{fig:fs}\includegraphics[width=0.4\linewidth]{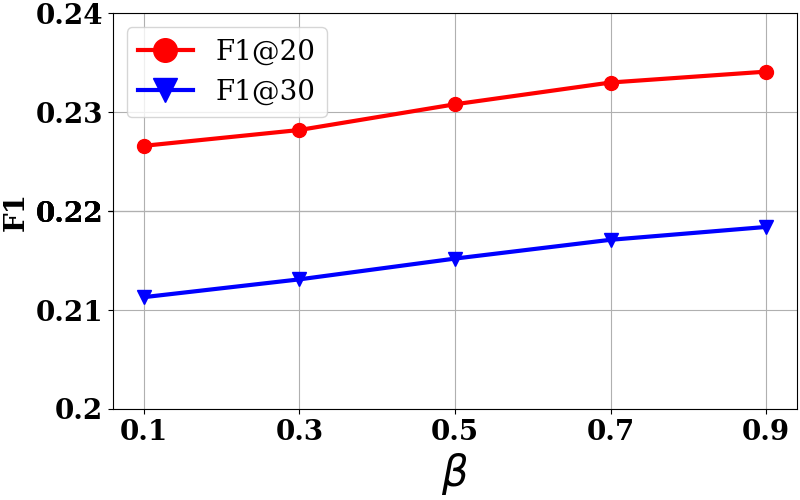}}
 \vskip -0.15in
\caption{The fairness and utility w.r.t different $\beta$.}
\label{fig:hyperparam_fig2}
\vskip -0.15in
\end{figure}

\subsection{Performance Comparison of Recommender Systems}\label{sec:comparison}
In this part, We present different methods' performance (user-side recommendation utility and item group fairness) evaluated on \textbf{top-30} ranking results. 

\begin{table*}[t]
\centering
\caption{Fairness and recommendation utility of different methods on two backbone models over datasets with a binary-valued item attribute (evaluated on top-30 recommendation results). \textbf{RI} denotes the relative improvement of \mymodel{} over UNS. We highlight the best results (in bold) and the \underline{second best results} with underline.}
\label{tab: table_comparison2}
\resizebox{0.9\linewidth}{!}{
\begin{tabular}{llllllllllllll}
\hline
\hline
                          &                  & \multicolumn{6}{c}{\textbf{ML1M-2}}                                                                                                                 & \multicolumn{6}{c}{\textbf{Amazon-2}}                                                                                   \\ \hline
Backbone                  & Method           & {\begin{tabular}[c]{@{}l@{}}Recall-\\ Disp.@30\end{tabular}}    & {\begin{tabular}[c]{@{}l@{}}Recall-\\ Min.@30\end{tabular}}     & \multicolumn{1}{l|}{\begin{tabular}[c]{@{}l@{}}Recall-\\ Avg.@30\end{tabular}}     & N@30            & P@30            & \multicolumn{1}{l|}{R@30}            & {\begin{tabular}[c]{@{}l@{}}Recall-\\ Disp.@30\end{tabular}}    & {\begin{tabular}[c]{@{}l@{}}Recall-\\ Min.@30\end{tabular}}     & \multicolumn{1}{l|}{{\begin{tabular}[c]{@{}l@{}}Recall-\\ Avg.@30\end{tabular}}}           & N@30            & P@30            & R@30            \\ \hline
& UNS              & 0.2837                                  & 0.3142                             & \multicolumn{1}{c|}{0.4344}                & 0.3817                   & \underline{0.1364}             & \multicolumn{1}{c|}{0.5589}          & 0.6697                              & 0.0530                             & \multicolumn{1}{c|}{0.1606}          & \underline{0.0899}             & \underline{0.0169}             & 0.1904                   \\
                           & NNCF             & 0.1592                                  & 0.3344                             & \multicolumn{1}{c|}{0.3978}                & 0.3393                   & 0.1193                   & \multicolumn{1}{c|}{0.5137}          & 0.5012                              & \underline{0.0796}                             & \multicolumn{1}{c|}{0.1592}          & 0.0840                   & 0.0161                   & 0.1538                   \\
                           & DNS              & 0.2822                                  & 0.3261                             & \multicolumn{1}{c|}{0.4543}                & \textbf{0.4097}          & \textbf{0.1426}          & \multicolumn{1}{c|}{\textbf{0.5940}} & 0.5875                              & 0.0722                             & \multicolumn{1}{c|}{\textbf{0.1750}} & \textbf{0.0972}          & \textbf{0.0181}          & \textbf{0.2021}          \\
                           & FairStatic       & \underline{0.0705}                         & \underline{0.4242}                    & \multicolumn{1}{c|}{\underline{0.4564}}       & 0.3740                   & 0.1323                   & \multicolumn{1}{c|}{0.5503}          & 0.5604                              & 0.0705                             & \multicolumn{1}{c|}{0.1604}          & 0.0887                   & 0.0164                   & 0.1856                   \\
                           & Reg              & 0.1928                                  & 0.3567                             & \multicolumn{1}{c|}{0.4419}                & 0.3495                   & 0.1342                   & \multicolumn{1}{c|}{0.5515}          & 0.5305                              & 0.0765                             & \multicolumn{1}{c|}{0.1630}          & 0.0871                   & 0.0166                   & 0.1834                   \\ \cline{2-14} 
                           & DPR              & 0.0913                                  & 0.4088                             & \multicolumn{1}{c|}{0.4499}                & 0.3479                   & 0.1315                   & \multicolumn{1}{c|}{0.5286}          & \underline{0.4735}                        & 0.0713                       & \multicolumn{1}{c|}{0.1275}          & 0.0719                   & 0.0142                   & 0.1633                   \\
                           & FairGAN-1        & 0.3227                                  & 0.2807                             & \multicolumn{1}{c|}{0.4145}                & 0.3564                   & 0.1267                   & \multicolumn{1}{c|}{0.5136}          & 0.6338                              & 0.0598                             & \multicolumn{1}{c|}{0.1632}          & 0.0751                   & 0.0136                   & 0.1478                   \\ \cline{2-14} 
                           & \textbf{FairNeg} & \textbf{0.0221}                         & \textbf{0.4680}                    & \multicolumn{1}{c|}{\textbf{0.4786}} & \underline{0.3880}             & 0.1360                   & \multicolumn{1}{c|}{\underline{0.5720}}    & \textbf{0.4706}                     & \textbf{0.0886}                    & \multicolumn{1}{c|}{\underline{ 0.1674}}    & 0.0888                   & 0.0168                   & \underline{0.1917}             \\ \cline{2-14} 
\multirow{-9}{*}{MF}       & \textbf{RI}      & 92.21\%                                 & 48.95\%                            & \multicolumn{1}{c|}{10.17\%}               & 1.65\%                   & -0.29\%                  & \multicolumn{1}{c|}{2.34\%}          & 29.73\%                             & 67.17\%                            & \multicolumn{1}{c|}{4.27\%}          & -1.22\%                  & -0.59\%                  & 0.68\%                   \\ \hline
                           & UNS              & 0.1896                                  & 0.3831                             & \multicolumn{1}{c|}{0.4727}                & \underline{0.4097}             & \underline{0.1434}             & \multicolumn{1}{c|}{\underline{0.6043}}    & 0.6347                              & 0.0631                             & \multicolumn{1}{c|}{0.1728}          & \underline{0.0979}             & \underline{0.0180}             & \underline{0.2048}             \\
                           & NNCF             & 0.1613                                  & 0.3780                             & \multicolumn{1}{c|}{0.4507}                & 0.3914                   & 0.1353                   & \multicolumn{1}{c|}{0.5753}          & \underline{0.5292}                        & 0.0799                             & \multicolumn{1}{c|}{0.1697}          & 0.0943                   & 0.0173                   & 0.1985                   \\
                           & DNS              & 0.2283                                  & 0.3620                             & \multicolumn{1}{c|}{0.4691}                & \textbf{0.4138}          & \textbf{0.1444}          & \multicolumn{1}{c|}{\textbf{0.6047}} & 0.6269                              & 0.0661                             & \multicolumn{1}{c|}{\textbf{0.1773}} & \textbf{0.1018}          & \textbf{0.0185}          & \textbf{0.2086}          \\
                           & FairStatic       & \underline{0.0462}                            & \underline{0.4541}                       & \multicolumn{1}{c|}{\underline{0.4866}}          & 0.3845                   & 0.1334                   & \multicolumn{1}{c|}{0.5770}          & 0.5361                              & \underline{0.0819}                       & \multicolumn{1}{c|}{0.1766}          & 0.0963                   & 0.0180                   & 0.2029                   \\
                           & Reg              & 0.1472                                  & 0.4069                             & \multicolumn{1}{c|}{0.4771}                & 0.4030                   & 0.1424                   & \multicolumn{1}{c|}{0.5994}          & 0.5753                              & 0.0745                             & \multicolumn{1}{c|}{0.1755}          & 0.0967                   & 0.0180                   & 0.2027                   \\ \cline{2-14} 
                           & DPR              & 0.0945                                  & 0.4356                             & \multicolumn{1}{c|}{0.4811}                & 0.3961                   & 0.1407                   & \multicolumn{1}{c|}{0.5929}          & 0.5864                              & 0.0712                             & \multicolumn{1}{c|}{0.1721}          & 0.0960                   & 0.0178                   & 0.2009                   \\
                           & FairGAN-2        & 0.3749                                  & 0.2633                             & \multicolumn{1}{c|}{0.4212}                & 0.3802                   & 0.1311                   & \multicolumn{1}{c|}{0.5479}          & 0.6220                              & 0.0631                             & \multicolumn{1}{c|}{0.1670}          & 0.0746                   & 0.0138                   & 0.1496                   \\ \cline{2-14} 
                           & \textbf{FairNeg} & \textbf{0.0415} & \textbf{0.4704}                    & \multicolumn{1}{c|}{\textbf{0.4908}}       & 0.4061                   & 0.1420                   & \multicolumn{1}{c|}{0.5981}          & \textbf{0.4952}                     & \textbf{0.0893}                    & \multicolumn{1}{c|}{\underline{0.1769}}    & 0.0974                   & 0.0179                   & 0.2004                   \\ \cline{2-14} 
\multirow{-9}{*}{LightGCN} & \textbf{RI}      & 78.11\%                                 & 22.79\%                            & \multicolumn{1}{c|}{3.82\%}                & -0.88\%                  & -0.98\%                  & \multicolumn{1}{c|}{-1.03\%}         & 21.98\%                             & 41.52\%                            & \multicolumn{1}{c|}{2.40\%}          & -0.55\%                  & -0.78\%                  & -2.15\%                  \\ \hline
\end{tabular}
}
\end{table*}

\begin{table*}[h]
\centering
\caption{Fairness and recommendation utility of different methods on two backbone models over datasets with a multi-valued item attribute (evaluated on top-30 recommendation results). \textbf{RI} denotes the relative improvement of \mymodel{} over UNS. We highlight the best results (in bold) and the \underline{second best results} with underline.}
\label{tab: table_comparison3}
\resizebox{0.9\linewidth}{!}{
\begin{tabular}{llllllllllllll}
\hline
\hline
                          &                  & \multicolumn{6}{c}{\textbf{ML1M-4}}                                                                                                                 & \multicolumn{6}{c}{\textbf{Amazon-4}}                                                                                   \\ \hline
Backbone                  & Method           & {\begin{tabular}[c]{@{}l@{}}Recall-\\ Disp.@30\end{tabular}}    & {\begin{tabular}[c]{@{}l@{}}Recall-\\ Min.@30\end{tabular}}     & \multicolumn{1}{l|}{\begin{tabular}[c]{@{}l@{}}Recall-\\ Avg.@30\end{tabular}}     & N@30            & P@30            & \multicolumn{1}{l|}{R@30}            & {\begin{tabular}[c]{@{}l@{}}Recall-\\ Disp.@30\end{tabular}}    & {\begin{tabular}[c]{@{}l@{}}Recall-\\ Min.@30\end{tabular}}     & \multicolumn{1}{l|}{{\begin{tabular}[c]{@{}l@{}}Recall-\\ Avg.@30\end{tabular}}}           & N@30            & P@30            & R@30            \\ \hline

\multirow{8}{*}{MF}       & UNS              & 0.2296                              & 0.2632                             & \multicolumn{1}{c|}{0.3783}          & 0.3215                   & 0.1270                   & \multicolumn{1}{c|}{0.4666}          & 0.6330                              & 0.0360                             & \multicolumn{1}{c|}{0.1228}          & 0.0854                   & \underline{0.0283}             & \underline{0.1350}             \\
                          & NNCF             & 0.1479                              & 0.3082                             & \multicolumn{1}{c|}{0.3898}          & 0.3321                   & 0.1280                   & \multicolumn{1}{c|}{0.4845}          & 0.5053                              & 0.0470                             & \multicolumn{1}{c|}{0.1085}          & 0.0707                   & 0.0244                   & 0.1164                   \\
                          & DNS              & 0.2164                              & 0.2997                             & \multicolumn{1}{c|}{\textbf{0.4058}} & \textbf{0.3516}          & \textbf{0.1348}          & \multicolumn{1}{c|}{\textbf{0.5087}} & 0.6377                              & 0.0348                             & \multicolumn{1}{c|}{0.1237}          & \textbf{0.0874}          & \textbf{0.0284}          & \textbf{0.1390}          \\
                          & FairStatic       & 0.0920                              & 0.3430                             & \multicolumn{1}{c|}{0.3800}          & 0.3012                   & 0.1206                   & \multicolumn{1}{c|}{0.4507}          & \underline{0.5035}                        & \underline{0.0501}                       & \multicolumn{1}{c|}{0.1257}          & 0.0781                   & 0.0265                   & 0.1249                   \\ \cline{2-14} 
                          & DPR              & \underline{0.0539}                        & \underline{0.3576}                       & \multicolumn{1}{c|}{0.3767}          & 0.2907                   & 0.1212                   & \multicolumn{1}{c|}{0.4505}          & 0.5919                              & 0.0320                             & \multicolumn{1}{c|}{0.0956}          & 0.0603                   & 0.0213                   & 0.0961                   \\
                          & FairGAN-1        & 0.2679                              & 0.2577                             & \multicolumn{1}{c|}{0.3768}          & 0.3189                   & 0.1215                   & \multicolumn{1}{c|}{0.4594}          & 0.6171                              & 0.0371                             & \multicolumn{1}{c|}{\underline{0.1264}}    & \underline{0.0863}             & 0.0276                   & 0.1286                   \\ \cline{2-14} 
                          & \textbf{FairNeg} & \textbf{0.0253}                     & \textbf{0.3858}                    & \multicolumn{1}{c|}{\underline{0.4028}}    & \underline{ 0.3333}             & \underline{0.1295}             & \multicolumn{1}{c|}{\underline{0.4864}}    & \textbf{0.4816}                     & \textbf{0.0594}                    & \multicolumn{1}{c|}{\textbf{0.1298}} & 0.0832                   & 0.0279                   & 0.1339                   \\ \cline{2-14} 
                          & \textbf{RI}      & 88.98 \%                              & 46.58 \%                             & \multicolumn{1}{c|}{6.46\%}            & 3.67\%                     & 1.97\%                     & \multicolumn{1}{c|}{4.24\%}            & 23.92\%                               & 65.00\%                              & \multicolumn{1}{c|}{5.72\%}            & -2.58\%                    & -1.41\%                    & -0.81\%                    \\ \hline
\multirow{8}{*}{LightGCN} & UNS              & 0.1664                              & 0.3253                             & \multicolumn{1}{c|}{0.4157}          & \underline{ 0.3609}             & \underline{0.1371}             & \multicolumn{1}{c|}{\underline{0.5228}}    & 0.6357                              & 0.0377                             & \multicolumn{1}{c|}{0.1330}          & \underline{0.0925}             & \underline{0.0302}             & \underline{0.1463}             \\
                          & NNCF             & 0.1033                              & 0.3366                             & \multicolumn{1}{c|}{0.3958}          & 0.3414                   & 0.1291                   & \multicolumn{1}{c|}{0.4994}          & 0.5326                              & 0.0498                             & \multicolumn{1}{c|}{0.1301}          & 0.0871                   & 0.0288                   & 0.1371                   \\
                          & DNS              & 0.1653                              & 0.3261                             & \multicolumn{1}{c|}{\underline{0.4183}}    & \textbf{0.3667}          & \textbf{0.1379}          & \multicolumn{1}{c|}{\textbf{0.5274}} & 0.6368                              & 0.0377                             & \multicolumn{1}{c|}{\underline{0.1364}}          & \textbf{0.0942}          & \textbf{0.0310}          & \textbf{0.1491}          \\
                          & FairStatic       & 0.0761                              & \underline{0.3720}                       & \multicolumn{1}{c|}{0.4123}          & 0.3472                   & 0.1309                   & \multicolumn{1}{c|}{0.5076}          & \underline{0.5290}                        & \underline{0.0551}                    & \multicolumn{1}{c|}{0.1346} & 0.0886                   & 0.0295                   & 0.1394                   \\ \cline{2-14} 
                          & DPR              & \underline{0.0638}                        & 0.3709                             & \multicolumn{1}{c|}{0.4121}          & 0.3448                   & 0.1334                   & \multicolumn{1}{c|}{0.5061}          & 0.5759                              & 0.0413                             & \multicolumn{1}{c|}{0.1318}          & 0.0887                   & 0.0290                   & 0.1397                   \\
                          & FairGAN-2        & 0.2982                              & 0.2408                             & \multicolumn{1}{c|}{0.3850}          & 0.3353                   & 0.1234                   & \multicolumn{1}{c|}{0.4843}          & 0.5402                              & 0.0454                             & \multicolumn{1}{c|}{0.1288}          & 0.0841                   & 0.0276                   & 0.1279                   \\ \cline{2-14} 
                          & \textbf{FairNeg} & \textbf{0.0280}                     & \textbf{0.4020}                    & \multicolumn{1}{c|}{\textbf{0.4184}} & 0.3545                   & 0.1338                   & \multicolumn{1}{c|}{0.5169}          & \textbf{0.4787}                     & \textbf{0.0556}                       & \multicolumn{1}{c|}{{\textbf{0.1411}}}    & 0.0905                   & 0.0295                   & 0.1418                   \\ \cline{2-14} 
                          & \textbf{RI}      & 83.17\%                               & 23.58\%                              & \multicolumn{1}{c|}{0.66\%}            & -1.77\%                    & -2.41\%                    & \multicolumn{1}{c|}{-1.13\%}           & 24.70\%                               & 46.15\%                              & \multicolumn{1}{c|}{1.15\%}            & -2.16\%                    & -2.35\%                    & -3.08\%                    \\ \hline
\end{tabular}
}
\end{table*}
    
\subsection{Ablation Study}\label{sec:ablation study}
In this part, we show the ablation study results on the Amazon-4 dataset, as presented in Figure~\ref{fig:ab_fig2}. The phenomenon is aligned with that on the ML1M-4 dataset. The incorporation of the importance-aware sampling distribution can help to maintain or even improve the overall utility, and the momentum and dynamic mechanism are both essential for item group fairness optimization.

\subsection{Hyper-parameter Analysis of \mymodel{}}\label{sec:hyp_exp}
In this part, we analyze the influence of two critical hyper-parameters in \mymodel{}: (i) the momentum parameter $\gamma$ in Equation~\ref{eq:grp_prob_2}; (ii) the fairness-utility tradeoff parameter $\beta$ in Equation~\ref{eq:beta}. For brevity, we report the results on the ML1M-4 dataset using LightGCN as the backbone, and similar patterns can be observed on other datasets/model backbones. \\
\textbf{Impact of $\gamma$.} To evaluate the influence of the momentum parameter, we vary its value in [0, 0.2] with step 0.05. The fairness (Recall-Disp) and recommendation utility (F1) results are evaluated under $K$ = 20 and 30 in Figure~\ref{fig:hyperparam_fig1}. Here, the importance-aware probability mixup ratio $\beta$ is set as 0, since we aim to investigate the momentum parameter's influence independently.
From the figures, we can find that the fairness performance improves under a proper momentum term (i.e., $\gamma = 0.1$ ) without hurting the utility.\\  
\textbf{Impact of $\beta$.} We also vary the fairness-utility tradeoff parameter in the negative sampling distribution mixup module between [0.1, 0.9] with step 0.2 and plot the results in Figure~\ref{fig:hyperparam_fig2}. The left figure indicates that with larger $\beta$, the fairness can be improved at first but then become worse under $K$ = 20 and 30. We hereby conclude that only properly incorporating importance-aware negative sampling advocates for a better utility-fairness tradeoff. The right figure indicates that by increasing the ratio of importance-aware sampling probability in the mixup procedure, the overall utility can be consistently improved, which attributes to the improvement of the negative samples' informativeness.

\end{document}